\documentclass[11pt]{article}
\usepackage{amsmath,amssymb}
\usepackage[dvips]{graphicx}

\def\Kp{$K^+$}
\def\Ko{$K^0$}
\def\Km{$K^-$}
\def\Kbo{$\bar K^0$}
\def\SigKN{ \mbox{$\Sigma_{KN}$} }
\def\sigm{\mbox{$\langle \sigma \rangle$}}
\def\psiN{\mbox{$\psi_N$}}
\def\LbN{$\Lambda$-$N$}

\def\psiL{\mbox{$\psi_\Lambda$}}
\def\psibN{\mbox{${\bar \psi}_N$}}
\def\psibL{\mbox{${\bar \psi}_\Lambda$}}
\def\gs{\mbox{$g_\sigma$}}
\def\gv{\mbox{$g_\omega$}}

\def\mKsq{\mbox{$m_K^2$}}
\def\gsL{\mbox{$g_\sigma^\Lambda$}}
\def\gvL{\mbox{$g_\omega^\Lambda$}}
\def\deLN{\mbox{$\Delta \varepsilon_{\Lambda N}$}}

\newcommand{\psla}{\mbox{\ooalign{\hfil/\hfil\crcr$p$}}}

\newcommand{\qsla}{\mbox{\ooalign{\hfil/\hfil\crcr$q$}}}

\newcommand{\vp}{\mbox{\boldmath $p$}}
\newcommand{\vq}{\mbox{\boldmath $q$}}

\newcommand{\epsi}{\mbox{$\varepsilon$}}

\newcommand{\tpsp}{\hspace{1.5em}}

\begin{document}
%\draft
\title{\Large \bf  Kaon Condensation and Lambda-Nucleon Loop
in the Relativistic Mean-Field Approach }

\author{Tomoyuki Maruyama$^{a,b,c}$ 
Takumi Muto$^d$, 
Toshitaka Tatsumi$^{e}$,\\
Kazuo Tsushima$^{f,g,h}$
and Anthony W. Thomas$^{i}$ \\
$^a$ Institute for Nuclear Theory,
University of Washington,
Seattle, WA 98195, USA \\
$^b$ College of Bioresource Science, Nihon University,
Fujisawa 252-8510, Japan\\
$^{c}$ Japan Atomic Energy Research Institute,
Tokai 319-1195, Japan \\
$^d$Chiba Institute of Technology, Narashino, Chiba 275-0023, Japan\\
$^e$Department of Physics, Kyoto University, Kyoto 606-8502, Japan\\
$^f$Instituto de F\'{\i}sica Te\'orica - UNESP, 01405-900, Sao Paulo, Brazil\\ 
$^g$Mackenzie University - FCBEE, 01302-907, Sao Paulo, Brazil\\ 
$^h$National Center for Theoretical Sciences at Taipei, Taipei 10617, Taiwan \\
$^i$Jefferson Laboratory, 12000 Jefferson Ave., Newport News,
VA 23606 USA 
}

%\date{January 29, 2005}

\maketitle

\bigskip

\begin{abstract}
The possibility of kaon condensation in high-density symmetric nuclear matter is investigated including both $s$- and $p$-wave kaon-baryon interactions within the relativistic mean-field (RMF) theory.
Above a certain density, we have a collective $\bar K_s$ state carrying the same quantum numbers as the antikaon. The appearance of the $\bar K_s$ state is caused by the time component of the axial-vector interaction between kaons and baryons. It is shown that the system becomes unstable 
 with respect to condensation of $K$-$\bar K_s$ pairs. We consider how the effective baryon masses affect the kaon self-energy coming from the time component of the axial-vector interaction.  Also, the role of the spatial component of the axial-vector interaction on the possible existence of the collective kaonic states is discussed in connection with $\Lambda$-mixing effects in the ground state of high-density matter. 
  Implications of $K\bar K_s$ condensation for high-energy heavy-ion collisions are briefly mentioned. 
\end{abstract}

\bigskip

%\pacs{PACS numbers: 13,75.Jz, 13,75.Ev, 21.65.+f, 21.80.+a, 24.10.Jv, 25.75.+r}

%\noindent
%Key Words: strage hadronic matter, relativistic mean-field theory,
%kaon condensation

\newpage

\section{Introduction}

\tpsp
The modification of kaon properties in high-density/temperature 
hadronic matter has been attracting much interest 
in the field of hadron physics, 
including the strangeness degree of freedom in heavy-ion collisions, 
especially in relation to kaon and/or antikaon production.  
Specifically, since the possible existence of antikaon condensation was 
suggested in neutron stars \cite{kn86}, 
its implications for astrophysical phenomena have been widely
discussed \cite{lbm95,t95,l96,pb97}. 
The examples are effects of softening of the equation of state on the
static and dynamic properties of neutron
stars \cite{tpl94,K-con,fmmt96,g01,bb94,p00,ty99} and 
thermal evolution of neutron stars through extra mechanisms 
of neutrino emission \cite{bkpp88,t88,fmtt94,pb90,t98}. 

It has been shown that the $s$-wave antikaon-nucleon ($\bar K N$)
attractive interactions given by the scalar interaction simulated by the
$KN$ sigma term ($\Sigma_{KN}$) and the vector interaction corresponding
to the Tomozawa-Weinberg term work as the main driving forces for kaon condensation. 
The same conclusion was obtained with scalar and vector interactions 
in the quark-meson coupling (QMC) model, but based on quark degrees of freedom\cite{tstw98}. 
 In neutron-star matter, the lowest antikaon excitation
energy, $\omega_{\rm min}$, decreases because of these $s$-wave $\bar KN$
interactions. 
At a certain density, $\omega_{\rm min}$ meets the antikaon chemical 
potential, which is equal to the electron chemical potential 
through the chemical equilibrium condition 
with respect to the weak interaction processes, 
$nn\rightleftharpoons npK^-$ and $nn\rightleftharpoons npe^-$. 
At this density, the system becomes unstable with respect to 
Bose-Einstein condensation (BEC) of the antikaons 
through the weak interaction processes, $nn\rightarrow npK^-$, 
$Ne^-\rightarrow NK^-\nu_e$\cite{mti00}. 
As a result, net strangeness is abundant 
in the fully-developed $K^-$-condensed phase.

On the other hand, in relativistic heavy-ion collisions, 
it is expected that many species of hadrons, in particular, 
a lot of strange particles, are produced in the hot and dense zone, 
where total strangeness is conserved because the typical time scales 
of the strong interaction are $O(10^{-23})$ sec, 
being much shorter than those of the strangeness-changing weak
interaction. 
In such {\it a strangeness-conserving system}, 
hyperons and kaons are produced  through the following typical
reactions, 
$NN\rightarrow NNK\bar K$, $NN\rightarrow N\Lambda K$. 
It has been suggested by Nelson and Kaplan \cite{Kprcon} 
that \Kp\Km(\Ko\Kbo)-pair condensation would occur 
in relativistic heavy-ion collisions. 
As the density increases, the lowest energy of \Kp(\Ko) is reduced by
the s-wave $KN$ interaction (mainly by the $KN$-sigma term, \SigKN) 
and the energy of \Kp(\Ko) eventually becomes equal to 
the strangeness chemical potential $\mu_s(=\mu_K)$, 
while that of \Km(\Kbo) reaches $-\mu_K$ simultaneously, 
and \Kp\Km(\Ko\Kbo)-pair condensation occurs without loss of energy. 
However, their conclusion seems to be unlikely in light of the subsequent
studies of the kaon properties in medium; the \Kp excitation energy 
acquires repulsion instead of attraction~\cite{h99,z04}.
Moreover, only nucleon matter was considered in Ref.~\cite{Kprcon} 
but no hyperon degrees of freedom was taken into account. 

Concerning strange particle production, hyperon and kaon pairs
are expected to be produced more favorably than kaon-antikaon pairs because of the different threshold energy:
the production energy of the lambda and kaon pair might be 
smaller than that of kaon and antikaon pair. 
In the case of lambda-kaon pair production, the strange baryon chemical
potential increases as the baryon density increases, and so the
strangeness chemical potential, $\mu_K$, has to increase so as to ensure
strangeness conservation for the system. 
When the lowest excitation energy of the $K^+$ becomes equal to $\mu_K$,
BEC occurs. Following this mechanism, the possibility of kaon ($K^+$)
condensation in relativistic heavy-ion collisions has been discussed
within hadronic models\cite{cle}. 

Triggered by the studies of kaon condensation, the kaon-nucleon
interaction in nuclei, in particular the antikaon optical potential, has
been elaborated in connection with kaonic atoms\cite{fgb94,fgm99}, 
subthreshold kaon and antikaon production in heavy-ion collisions 
and/or proton-nucleus collisions\cite{b97,l99,ss99}. 
The nature of the $\Lambda$(1405) in nuclear matter 
has been discussed in relation to the $s$-wave antikaon-nucleon
scattering amplitude\cite{k94,l98}. 
In line with their studies, theoretical work has been carried out on the
antikaon optical potential with coupled channel approaches based on
chiral models\cite{wkw96,ro00,ske00}, or 
with a reaction matrix method\cite{trp02}. 
There is still some controversy about the magnitude of 
the depth of the antikaon optical potential\cite{oor00}. 
On the assumption that the $\Lambda$(1405) is a bound state 
of the antikaon and nucleon due to the strongly attractive
antikaon-nucleon interaction, deeply bound kaonic nuclei have been 
recently proposed\cite{ay02,d04,k99}, and some experimental evidence
has been reported\cite{i04}. 

For the kaons, it has been shown that the modification influences 
kaon production in heavy-ion collisions \cite{f94,lkl95,tst00}, 
the dilepton production rate in the Relativistic
Boltzmann-Uhling-Uhlenbeck approach (RBUU)~\cite{LiKo} and/or the
structure of the fireball~\cite{Tatsu}. 

Meanwhile, recent hypernuclear experiments have opened a new stage to explore the hyperon-nucleon and hyperon-hyperon interactions\cite{gh95}.   
The information obtained in these studies of in-medium kaon dynamics in nuclei and hypernuclear physics may in turn provide us with a clue to clarify the interrelations between kaons and hyperons and the existence of kaon/antikaon condensations formed in  heavy-ion collisions and/or in neutron stars. 
In case of neutron stars, hyperon-mixing in a dense matter has been discussed in a realistic situation\cite{ell,sch,hyp1,hyp2,hyp3}. The interplay between kaons and hyperons has been considered by several authors\cite{ell,sch,m93,kvk95,bb01,m02,kv03}. In particular, mechanisms for kaon condensation caused by the $p$-wave kaon-baryon interaction have been elucidated\cite{m93,kvk95,m02,kv03}. 

In order to study the high-density hadronic matter realized in heavy-ion collisions, relativistic  covariant kinematics is necessary. A plausible approach is the relativistic mean-field (RMF) theory, where meson-baryon interactions are naturally incorporated in the relativistic framework.  
The possibility of producing strange hadronic matter in
relativistic heavy-ion collisions was suggested on the basis of the RMF
approach, where the abundance of $\Lambda$ or other hyperons overwhelms
that of the nucleons~\cite{Rufa,sg00}. 
However, only baryons were considered in Ref.~\cite{Rufa,sg00}
rather than kaons, which are the lightest mesons  carrying strangeness. 
Because the properties of kaons are expected to be 
modified significantly by the surrounding nuclear medium, 
it is necessary to take them into account properly  
to explore the strange hadronic matter as well as hyperons.

In previous work, two of the authors (T.~Maruyama and T. T.) studied~\cite{kpcon,shin} strange hadronic matter by extending the relativistic mean-field (RMF) theory 
to incorporate nucleons, $\Lambda$ hyperons and
kaons, which are essential degrees of freedom.
These authors showed the possibility of $K^{+}(K^0)$-condensation
in high-density nuclear matter. 
When it occurs, the nucleon decays to a kaon-lambda ($K$-$\Lambda$) 
pair spontaneously. 
As a relativistic effect, it has been shown in Refs.~\cite{K-con,fmmt96} 
that the energy gain due to kaon condensation is moderated 
in neutron-star matter by the $self$-$suppression$ mechanism, 
which is caused by the suppression of the $s$-wave scalar interaction 
simulated by the $KN$-sigma term (\SigKN), 
due to the small effective nucleon mass. 
This self-suppression effect should also reveal itself 
in symmetric nuclear matter.

In this paper,  we extend the formalism of Ref.~\cite{kpcon}
to incorporate the lambda-nucleon (\LbN) loop into the kaon self-energy,
and study kaon properties in dense symmetric nuclear matter. 
First, the solutions to the dispersion equation for the kaon are
obtained by including the real part of the kaon self-energy. 
We find two collective modes in addition to the usual kaon and 
antikaon ones. 
However, we show that one of the collective modes, 
which lies within the continuum region of the $\Lambda$ particle-nucleon 
hole excitation, does not really exist on the physical sheet, 
by analyzing the dispersion equation in the complex energy plane.\footnote{Detailed analysis of the pion dispersion
equation in the complex energy plane has also been done in
Ref.~\cite{s99}.}   
We discuss in detail the effects of the $p$-wave
kaon-baryon interaction of the axial-vector coupling on the kaon
dispersion relations and clarify the onset mechanisms of kaon
condensation in the relativistic framework. 

The paper is organized as follows. In Sec. 2, we present the formalism for the expression of the $\Lambda$-$N$ loop contribution to the kaon self-energy in the RMF theory. Section 3 gives numerical results and related discussion. In Sec.~4, we discuss the effects of the baryon masses on the  $\Lambda$-$N$ loop contribution to the kaon self-energy coming from the 
time component of the axial-vector coupling. 
Hyperon-mixing effects are also addressed with reference to kaon condensation 
in neutron-star matter. Section 5 is devoted to a summary and concluding remarks. 
In the Appendix, we analyze the existence of the collective kaonic states by considering the solution to the dispersion equation in the complex kaon energy plane. 

\section{Formalism}

\tpsp
First, we briefly explain the basic formalism. We use the Lagrangian density: 
\begin{eqnarray}
{\cal L} & = & 
{\psibN} (i{\not \partial} - M_N )\psiN 
+ \gs {\psibN} \psiN \sigma
+ \gv {\psibN} \gamma_\mu \psiN \omega^\mu
\nonumber \\
&&+ \psibL (i{\not \partial} - M_{\Lambda} )\psiL 
+ \gsL \psibL \psiL  \sigma
+ \gvL \psibL \gamma_\mu \psiL \omega^\mu
- {\widetilde U} [\sigma] 
+ \frac{1}{2} m_{\omega}^2 \omega_\mu \omega^\mu
\nonumber \\
&& + \Big\lbrace\Big( {\partial_\mu} - \frac{3i}{8 f^2} {\psibN} \gamma_\mu  \psiN \Big)
{\phi_K^\dagger} \Big\rbrace  
\Big\lbrace \Big( {\partial^\mu} + \frac{3i}{8 f^2} {\psibN} \gamma^\mu  \psiN \Big)\phi_K \Big\rbrace 
\nonumber \\
&&- {\mKsq} {\phi_K^\dagger} \phi_K
+ \frac{\SigKN}{f^2} {\psibN} \psiN \phi_K^{\dagger} \phi_K \nonumber \\
&&+ i\widetilde f{\psibL} \gamma_{5} \gamma^{\mu} \psiN 
\partial_{\mu} \phi_{K} ~+~ {\rm h.c.} ,
\label{Lagrangian}
\end{eqnarray}
where $\psiN$, $\psiL$, $\sigma$, $\omega$ and $\phi_K$ are 
the nucleon, lambda, sigma-meson, omega-meson and kaon fields, 
respectively, ${\widetilde U} [\sigma]$ is the self-energy potential 
of the scalar mean-field given in Refs. \cite{K-con,TOMO2}, $f$
($\simeq$ 93 MeV) the meson decay constant, and $m_K$ is the kaon mass.  
In Eq.(\ref{Lagrangian}), isospin-dependent interaction terms are
suppressed, except for the $KN$ interaction, 
because we will treat only the isospin symmetric system in this study. 
Furthermore, we assume that the kaon-nucleon interaction is given solely by 
the $KN$~sigma term, \SigKN, and the Tomozawa-Weinberg (TW) type vector interaction.
The latter interaction is introduced in the same scheme as  the
''minimal'' coupling between the kaon current and vector mesons 
in the one-boson-exchange (OBE) model\cite{sch}.\footnote{This 
coupling scheme is different from that used in the
standard chiral perturbation theory\cite{kn86,lbm95,t95,l96}. 
Nevertheless, the quantitative difference is small 
between the two coupling schemes. }
The last term in Eq.~(\ref{Lagrangian}) is the $p$-wave kaon-baryon
interaction 
with the axial-vector coupling 
with $\widetilde f$($\equiv f_{KN\Lambda}/m_K$) being the $KN\Lambda$ coupling constant divided by the kaon mass. We take $\widetilde f=(D+3F)/(2\sqrt{3}f) \simeq 0.61/ f$, with $D$(=0.81) and $F$ (=0.44) being the coefficients of the axial-vector coupling terms for the kaon-baryon interaction in the effective chiral Lagrangian\cite{kn86,m93}. 

Then, the propagators for nucleon and lambda are given by 
\begin{eqnarray}
S_N (p) &=& ( \psla^* + M_N^{*}) 
\left[\frac{1}{p^{*2} - M_N^{*2} + i \delta} +
 \frac{i \pi}{E_N^* (\vp)} n(\vp) \delta(p_0 - \epsi_N (\vp) ) \right], 
\label{probN}
 \\
S_\Lambda (p) &=& 
\frac{\psla^* + M^*_\Lambda }
{p^{*2} - M_\Lambda^{*2} + i \delta}, 
\label{probL}
\end{eqnarray}
respectively, where
$M^*_\alpha$ ($\alpha = N, \Lambda$) and $p^*$ are the effective mass
and momentum of the corresponding particles defined by 
\begin{eqnarray}
M^*_{N(\Lambda)} &=& M_{N(\Lambda)} - U_s (N(\Lambda)) ,
\\
p^{*}_\mu &=& p_\mu - U_0 (N(\Lambda)) \delta^0_\mu ,
\\
E_{N(\Lambda)}^*(\vp) &=& \sqrt{\vp^2 + M^{*2}_{N(\Lambda)}} ,
\\
\epsi_{N(\Lambda)}(\vp) &=& E_{N(\Lambda)}^*(\vp) +
 U_0 (N(\Lambda)).  
\end{eqnarray}
In the above, 
$U_s$ and $U_0$ are the scalar and vector self-energies,
respectively, given by 
\begin{eqnarray}
U_{s}(N(\Lambda)) & = & \gs (\gsL) \sigm ,
\\
U_{0} (N(\Lambda))  
& = & \frac{g^2_\omega (g_\omega g^{\Lambda}_\omega) }{m_{\omega}^2} \rho_N  \ , 
\end{eqnarray}
with $\rho_N$ being the nucleon number density. 

Now we consider the kaon self-energy arising from the \LbN~loop.
First, we define the propagator of the kaon with momentum 
$q=(q_0; \vq)$ as
\begin{eqnarray}
D_K^{-1} (q) &=& \Delta_s^{-1} (q) - \Pi_K (q) ,
\nonumber \\
&=& ( q_0 - U_{0}(K) )^2 - \vq^2 - m_K^{*2} - \Pi_K (q) , 
\label{DKinverse}
\end{eqnarray}
with 
\begin{eqnarray}
m_K^{*2} &=& \mKsq - \SigKN \rho_s (N)/f^2 , 
\label{Kefm}
\\
U_{0}(K) &=& \frac{3}{8 f^2} {\rho_N}, 
\end{eqnarray}
where 
$\rho_s(N)$ is the nucleon scalar density. 
In Eq.(\ref{DKinverse}), $\Pi_K$ stands for the kaon self-energy contribution  
from the \LbN~loop, which is given by 
\begin{equation}
\Pi_K (q) = i {\widetilde f}\ ^2 \int \frac{d^4 p}{(2 \pi)^4}
{\rm Tr} \{ \qsla \gamma_5 S_{\Lambda} (p-q) \gamma_5 \qsla S_{N} (p) \} .
\label{CorK}
\end{equation}
Substituting Eqs.(\ref{probN}) and (\ref{probL}) into 
Eq.(\ref{CorK}), one obtains the $\Lambda$ particle-nucleon hole ($\Lambda N^{-1}$) contribution\footnote{The superscript ``-1'' denotes a hole state. }  to the kaon self-energy: 
\begin{eqnarray}
\Pi_K (q) &=&  - \frac{1}{4 \pi^3} {\widetilde f}\ ^2
\int \frac{d^3 p}{E_N^* (\vp)} n(\vp)
\nonumber \\
& & ~~~~~ \times
 \left.\frac{2 (p \cdot q)^2 - 2 (p \cdot q) (q^* \cdot q)
+ q^2 (p \cdot q^\ast - M_N^{*2} - M_N^{*} M_{\Lambda}^{*}  )}
{ 2 q^* \cdot p - q^{*2} - M_N^{*2} + M_\Lambda^{*2} - i \delta }
\right|_{p_0 = E_N^{*}(\vp)} ,
\label{eq:self}
\end{eqnarray}
where 
\begin{eqnarray}
q^*_\mu &=& q_\mu + \Delta U_0 \delta^{0}_\mu ,
\\
\Delta U_0 &=& U_0 (\Lambda) - U_0 (N) .
\end{eqnarray} 

For later discussions, 
we give also the kaon self-energy at $Q\equiv |{\bf q}|=0$: 
\begin{equation}
\Pi_K (q_0; Q=0) =
\frac{1}{\pi^2}{\widetilde f}\ ^2 q_0^2
\int \frac{dp p^2\, n(\vp)}{E_N^* (\vp)} 
\frac{  2 E_N^{*2} (\vp) - q_0^* E_N^*(\vp)
- M_N^* (M_N^* + M_{\Lambda}^* ) }
{q_0^{*2} - 2 q_0^* E_N^*(\vp) + M_N^{*2} - M_\Lambda^{*2} + i \delta } \ . 
\label{eq:Res0}
\end{equation}
Note that the kaon self-energy, $\Pi_K(q_0; {\bf 0})$, has non-zero imaginary parts 
in two energy regions:
$\epsi_N (0) -  \epsi_\Lambda (0) < q_0 < \epsi_N (p_F(N)) -  \epsi_\Lambda (p_F(N))$
and   
$\epsi_N (0) +  \epsi_\Lambda (0) < q_0 < \epsi_N (p_F(N)) +  \epsi_\Lambda (p_F(N))$ (with  $p_F(N)$ the Fermi momentum of the nucleon), 
because the integrand in Eq.(\ref{eq:Res0}) diverges at $p=p_c$ with
\begin{equation}
p_c = \sqrt{ \frac{[(q_0^* - M_N^*)^2 - M_{\Lambda}^{*2})]
                  [(q_0^* + M_N^*)^2 - M_{\Lambda}^{*2})]}
{4 q_0^2} } .
\end{equation}
These two regions correspond to the $\Lambda$ particle - nucleon hole
($\Lambda N^{-1}$) and anti-$\Lambda$ - nucleon hole ($\bar\Lambda
N^{-1}$) continuum states. Accordingly, the
above continuum states appear as branch cuts of $D_K^{-1} (q_0;\vq)$ in the complex $q_0$ plane. 
In the actual calculation
we define energies of kaonic states as pole energies,   
${\tilde \omega}$, of ${\rm Re} D_K^{-1} (q_0;\vq)$, 
${\rm Re} D_K^{-1} (q_0 = {\tilde \omega};\vq) = 0$. 

It is to be noted that the second-order perturbation with respect to 
the axial-vector current (the second-order effect, abbreviated as SOE)
 which leads to the same order as that of the KN sigma term [$O(m_K^2)$ ], 
is necessary in order to reproduce the on-shell $s$-wave $K$ ($\bar K$)-$N$ 
scattering lengths\cite{lbm95,t95,l96,fmmt96,brown94}. 
The SOE consists of the smooth part proportional to $q_0^2$ and the pole part from the $\Lambda$(1405). 
In matter, it has been shown that the SOE becomes negligible for the $K^-$ self-energy 
owing to the decrease in the $K^-$ energy, 
while it works repulsively for the $K^+$ with a sizable increase 
in the $K^+$ energy from the free kaon mass. 
However, as we see in Sec.~\ref{subsec:cond}, the repulsive effect on the $K^+$ is 
reproduced (at least qualitatively) without the SOE in the relativistic framework, 
as a result of the self-suppression mechanism for the $s$-wave $KN$ scalar 
interaction\cite{K-con,fmmt96}. 
Therefore, throughout this paper, we omit the SOE and consider the simplified expression for the kaon self-energy in nuclear matter. 

\section{Numerical results}
\label{sec:results}

\subsection{Parameters}
\label{subsec:param}

\tpsp
We use the parameter sets of PM1 \cite{K-con}, which give  
the binding energy per baryon $BE = 16$ MeV, $M_N^*/M_N=0.7$ at $\rho_0 $= 0.17 fm$^{-3}$. 
The $KN$ sigma term has some ambiguities, and its value is taken to be
$\SigKN$ = 400 MeV here, as an example\cite{dll96}. 
We consider only the matter with no net strangeness, and  parameterize the scalar and vector (time component) parts for the $\Lambda$-self-energy as
\begin{eqnarray}
U_s (\Lambda) = c_s U_s (N) &~~{\rm and}~~&
U_0 (\Lambda) = c_0 U_0 (N) \ ,
\end{eqnarray}
respectively. 
Following Ref.~\cite{kpcon}, we adopt two parameter sets for the
coupling of $\Lambda$ to the nucleon mean-fields. One set (L1) is 
that $c_s = c_0 = 2/3$ based on the $SU(6)$  symmetry relation \cite{DoverGal}, 
and the other set (L2) is that $c_0 = 0.17$ 
(which is the minimum value given in Ref. \cite{Rufa}), and 
$c_s$ is obtained by the following relation, 
\begin{equation}
U_s(\Lambda) - U_0(\Lambda) = \frac{2}{3} ( U_s(N) - U_0(N) ).
\label{ucent}
\end{equation}
A similar relation is found naturally within the QMC model\cite{tsht98}. 

\subsection{Condensation of $K$-$\bar K_s$ pairs with null momentum}
\label{subsec:cond}

\begin{figure}[ht]
\vspace{0.5cm}
\hspace*{1cm}
{\includegraphics[scale=0.8]{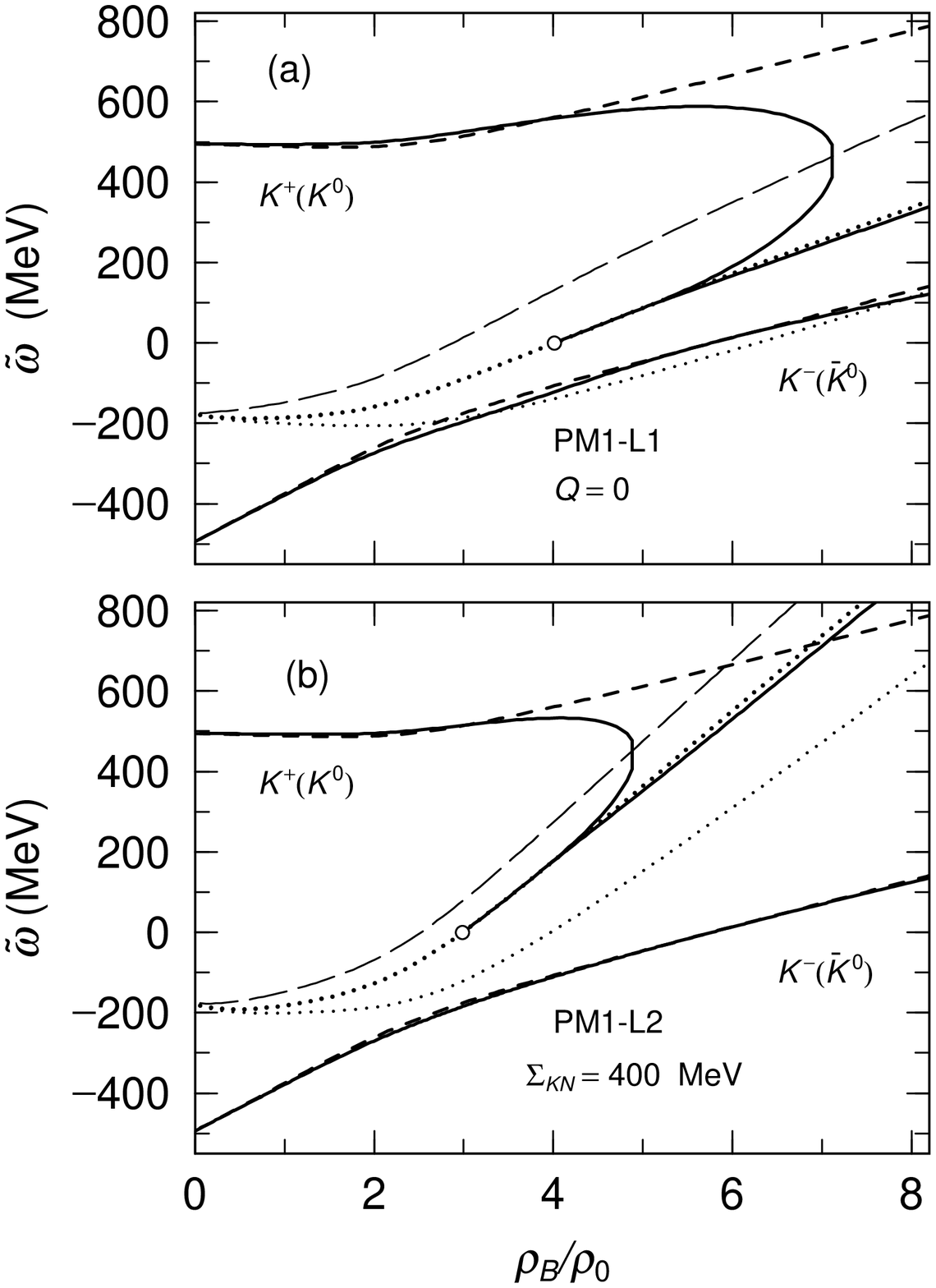}}
\caption
{\small 
Density-dependence of the pole energies of the kaon propagator 
with zero momentum $\tilde \omega$
using the parameter-sets PM1-L1 (a) and PM1-L2 (b). 
The solid and dashed lines show ${\tilde \omega}$ with 
the \LbN~loop and without it, respectively.
The open circles denote the points at ${\tilde \omega} = 0$. 
The meanings of the dotted [thin-dotted] \{thin-long-dashed\} lines
are explained in the text.}
\label{Kmas}
\end{figure}

\tpsp
First, we consider the pole energy, ${\tilde \omega}$,  
of the kaon propagator with zero momentum \break
(${\rm Re}D_K^{-1}({\tilde \omega};{\bf 0})=0$). 
The effect of finite momentum, $Q=|{\bf q}|\neq 0$, 
will be discussed separately.
At $Q=0$, the self-energy from the $\Lambda$-$N$ loop comes only from
the time component of 
the axial-vector coupling term for the kaon-baryon interaction [ Eq.~(\ref{eq:Res0})] . 
In Fig.~\ref{Kmas}, we show the density-dependence of ${\tilde \omega}$
in the no-net-strangeness system ($\rho_\Lambda = \rho_K = 0$) using the two parameter sets 
PM1-L1 (a) and PM1-L2 (b).
The solid and dashed lines represent the results 
with and without the \LbN~loop, respectively.
For a reference, we also plot the density-dependence of minus 
the $\Lambda N^{-1}$ energy with zero momentum, 
$- \deLN (p_F(N)) = \epsi_{N}(p_F(N)) - \epsi_\Lambda (p_F(N))$ 
[$-\deLN(0) = \epsi_{N}(0) - \epsi_\Lambda (0)$] 
\{$\epsi_{N}(p_F(N)) - \epsi_\Lambda (0)$\} 
with the dotted [thin-dotted] \{thin-long-dashed\} line.

In case of no \LbN~loop, there are two branches 
corresponding to \Kp(\Ko) and \Km(\Kbo).
Noting that the energy of the antikaon branch \Km(\Kbo) should be read as $-\tilde\omega$,  
we denote the energy as $\omega_K(K) = {\tilde \omega}$ for the kaon, 
$K$,  and $\omega_K(\bar K) = - {\tilde \omega}$ for the antikaon, ${\bar K}$. 
The kaon energy increases monotonically as density increases, while the
antikaon energy decreases. As a result, the critical condition for 
$K^+K^-(K^0\bar K^0)$- pair condensation, $\omega_K(K)+\omega_K(\bar
K)=0$, 
proposed in Ref.~\cite{Kprcon} is never met at any baryon number density. 
In Ref.\cite{kpcon}, two of the authors instead proposed the \Kp (\Ko) condensation scenario, 
where the nucleon on the Fermi surface decays into the lambda and kaon
pair when the nucleon Fermi energy becomes larger than
the total energy of the lambda mass and the kaon mass. 
In Fig.~\ref{Kmas}, we can see that this condensation occurs in the density region 
where the dashed lines are smaller than the thin-long-dashed lines. 
The critical density of the \Kp (\Ko) condensation is   
about 16 $ \rho_0$ for PM1-L1 and 6 $ \rho_0$ 
for PM1-L2 parameter sets, respectively.
The former cannot be seen in Fig.~\ref{Kmas}a.

Introducing the $\Lambda N^{-1}$~loop contribution, however,
${\tilde \omega}$ 
exhibits a complicated density-dependence as shown by the solid lines in
Fig.~1. First, we can see that the \LbN~loop contributes very little to the 
\Km (\Kbo) energy in the whole density region.
As for the \Kp (\Ko) energy, this effect is also small in the low density
region, while it becomes pronounced as the density increases.

We can see two additional branches, denoted by $\bar K_s$
and $K_s$, simultaneously appearing above a certain density, $\rho_c^{(1)}$, 
which satisfies $\deLN (p_F(N)) = 0$. 
The density $\rho_c^{(1)}$  can be read as 4.0 $\rho_0$ for  PM1-L1 and 3.0 $\rho_0$ for PM1-L2. 
These branches have the same quantum numbers as antikaons and kaons, respectively, as shown below. 
As the density increases, the energy difference between $K_s$ and $\bar
K_s$ becomes larger and the $K^+(K^0)$ branch merges with the $\bar K_s$
branch at a density, $\rho_c^{(2)}$, and there only the $K^-(\bar K^0)$ and $K_s$ branches are left above that density. The density $\rho_c^{(2)}$ is read as 7.2 $\rho_0$ for  PM1-L1 and 4.9 $\rho_0$ for PM1-L2. 

\begin{figure}[ht]
\hspace*{0.3cm}
{\includegraphics[scale=0.7]{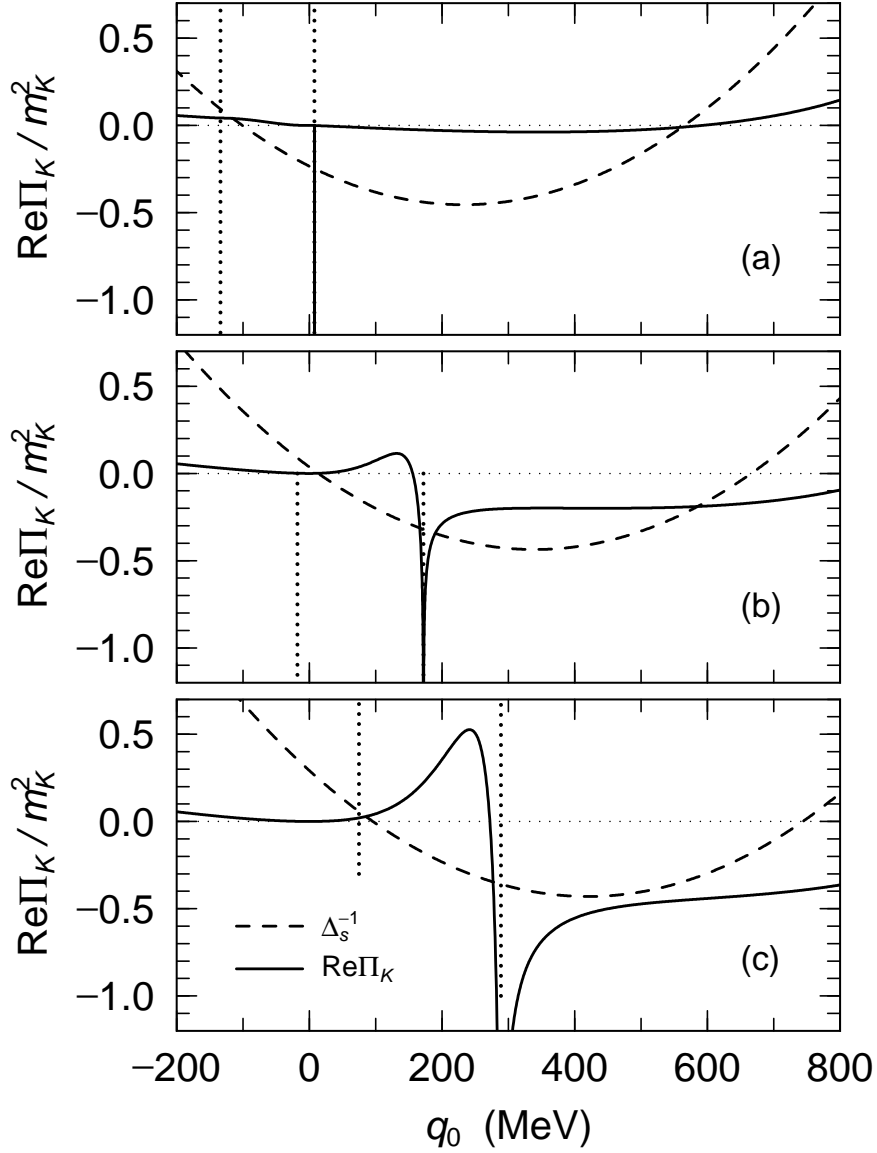}}
\caption{\small
The real part of the self-energy due to the \LbN~loop, 
${\rm Re} \Pi_K$ with zero momentum $Q = 0$ with the parameter set PM1-L1 for  
$\rho_B = 4.1 \rho_0$ (a), $6 \rho_0$ (b) and $8 \rho_0$ (c) (the solid lines) . 
The dashed lines represent $\Delta_s^{-1}(q_0;0) $. 
The dotted lines indicate the energies $q_0 = - \deLN(0)$ and
 $q_0 = - \deLN(p_F(N))$. }
\label{Kres}
\end{figure}

In order to analyze the results more in detail, we show in Fig.~\ref{Kres} 
the self-energy, ${\rm Re} \Pi_K(q_0;{\bf 0})$ (solid lines),  
and the inverse of kaon propagator without the $\Lambda-N$ loop, 
$\Delta_s^{-1} (q_0;{\bf 0})$ (dashed lines), as functions of $q_0$ 
with the parameter set PM1-L1 for  
$\rho_B $= 4.1 $\rho_0$ (a), 6 $\rho_0$ (b) and 8 $\rho_0$ (c). 
The dotted lines denote the boundaries of the $\Lambda N^{-1}$ continuum
region:  
$q_0 = - \deLN(0)$ for the lower boundary and 
 $q_0 = - \deLN(p_F(N))$ for the higher one. 
The energies $q_0$ at the intersection points of the solid and dashed lines
indicate the pole energies ${\tilde \omega}$ of ${\rm Re}D_K(q_0;{\bf 0})$. 
There are four intersection points of ${\rm Re} \Pi_K (q_0;{\bf 0})$ 
with $\Delta^{-1}_s (q_0; {\bf 0})$ in Fig.~\ref{Kres}b. The residue at the pole, $1/(\partial {\rm Re}D_K^{-1}/\partial q_0)_{q_0=\tilde\omega} $, is negative 
for the first and the third poles in the increasing order of $\tilde\omega$ and  positive 
for the second and the fourth poles.  From the sign of the residue for each pole, 
we can assign the strangeness of these states, $s = -1$, $+1$, $-1$
and $+1$ in increasing order of $\tilde\omega$, which are called ${\bar K}$,
$K_s$, ${\bar K}_s$, and $K$.
Note that
$\partial {\rm Re}D_K^{-1}/\partial q_0 \approx \partial \Delta_s^{-1}/\partial q_0$ 
for the first and fourth poles, while 
$\partial {\rm Re}D_K^{-1}/\partial q_0 \approx - \partial {\rm Re}\Pi_K/\partial q_0$ 
for the second and third poles.

Here we must mention that the above calculations have been done by the
use of the real part of the self-energy, ${\rm Re}\Pi_K(q_0;{\bf 0})$. If
these pole energies give the vanishing imaginary part, ${\rm
Im}\Pi_K(\tilde\omega;{\bf 0})=0$, they are mathematically verified and
physical modes. $\bar K, K, \bar K_s$ satisfy this criterion.
However, if it gives a finite imaginary part, ${\rm
Im}\Pi_K(\tilde\omega;{\bf 0})\neq 0$, we must carefully check whether
it has a physical meaning. Actually, the pole energy of $K_s$ gives a finite
imaginary part, because it lies in the $\Lambda N^{-1}$
continuum region. We can see that $K_s$ is an unphysical branch by
extending our analysis in the complex plane: 
directly solving the dispersion
equation, $D_K^{-1}(q_0;{\bf 0})=0$, in the complex $q_0$ plane, we cannot
find any pole corresponding to $K_s$ on the physical sheet. 
The situation is unchanged even in nonrelativistic kinematics. 
More detailed discussions are given in the Appendix.

We can examine the appearance and disappearance of the $\bar K_s$ and $K_s$ modes more carefully.
In Eq.(\ref{eq:Res0}) the integrand has a singular point,
at $q_0 = -\deLN(p)$.
As the kaon energy $q_0$ decreases, the momentum integral becomes 
divergent at the point, $q_0 = - \deLN (p_F)$, because 
${\rm Re} \Pi_K (q_0;{\bf 0}) \sim - q_0^2 \log|q_0 + \deLN (p_F(N))|$.
Then, there appears a collective pole of ${\rm Re}D_K(q_0;{\bf 0})$ ($\bar K_s$) in the region $q_0 > -\deLN (p_F(N))$. 
Further decreasing $q_0$, one finds that the singular point given by $q_0=-\deLN(p)$
lies  in the continuum region, i.e., $ - \deLN(0) < q_0 <
-\deLN(p_F(N))$, and the sign of the integrand changes when crossing the
singular point. Thus the sign of ${\rm Re}\Pi_K (q_0;{\bf 0})$ changes in the region $ -\deLN(0) < q_0 < -\deLN(p_F(N))$.
Subsequently another unphysical collective pole of ${\rm Re}D_K (q_0;{\bf 0})$ ($K_s$)  appears in this continuum region.

Note that ${\rm Re}\Pi_K(q_0;{\bf 0}) = 0$ at $q_0 = 0$ 
because of the pre-factor $q_0^2$ in Eq.(\ref{eq:Res0}), and that
there is no collective mode when  the minus $\Lambda N^{-1}$ energy 
$\deLN (p_F(N))=0$.
Furthermore, this factor also makes the divergent peak very narrow, and 
the difference between the two pole energies becomes very small 
when  the minus $\Lambda$-hole energy $|\deLN (p_F(N))| \ll 1$ (see Fig.~2a). 

In principle this divergence occurs at any density except
 $\rho_c^{(1)}$, where $q_0 = - \deLN(p_F(N))$, 
but the peak is too tight and cannot be obtained numerically
at lower density  $\rho_B < \rho_c^{(1)}$. 
This divergence is caused by the rigid Fermi surface.
If we take into account the high momentum tail in the momentum distribution
function $n (\vp)$, the minimum depth of the kaon self-energy ${\rm Re} \Pi_K (q_0; {\bf 0})$ becomes finite.
In such a low energy region,  $\rho_B < \rho_c^{(1)}$, 
the kaon self-energy does not have
even a peak, and the collective modes must disappear. 

From these results, we conclude that there exists only one  collective mode, $\bar K_s$, which is physically meaningful. 
As the density increases, the nucleon Fermi energy exceeds the kaon energy and 
the collective mode, ${\bar K}_s$, appears with zero energy at $\rho_c^{(1)}$, which  
is a signal for an instability of nuclear matter, ${\bar K}_s$
condensation. However, it carries strangeness $-1$ and we cannot
observe such an instability in nuclear matter due to the conservation of
strangeness in strong-interaction processes. 
With a further increase in density, the kaon ($K$) 
and the collective (${\bar K}_s$) branches merge 
at $\rho_B =\rho_c^{(2)}$, where the double-pole condition, ${\rm Re}
D_K^{-1}(q_0; {\bf 0})=0$ and $\partial {\rm Re} D_K^{-1}(q_0; {\bf 0})
/\partial q_0=0$, is satisfied. 
As already mentioned, the actual energies of the ${\bar K}$ and ${\bar
K}_s$ branches must have the opposite sign to the pole energies.
Thus, at this critical density $\rho_c^{(2)}$, the sum of the two
energies of the $K$ and ${\bar K}_s$ branches becomes zero, and  the $K$ and ${\bar K}_s$ pair spontaneously appears.  
This indicates a kind of kaon condensation, 
namely $K{\bar K}_s$ condensation. 
It is worth mentioning that this mechanism of pair condensation is 
similar to that of $p$-wave charged pion condensation 
in neutron-star matter\cite{bc79,ew88}. 
In the case of in-medium pion dispersion relations, 
the spin-isospin sound mode called $\pi_s^+$, 
carrying the quantum number of $\pi^+$, 
appears through the proton particle - neutron hole excitations 
induced by the $p$-wave $\pi N$ interaction. 
At a certain density, the system becomes unstable 
with respect to creation of $\pi^-$$\pi_s^+$ pairs. 

\begin{figure}[ht]
\vspace{0.5cm}
\hspace*{1cm}
{\includegraphics[scale=0.8,angle=-90]{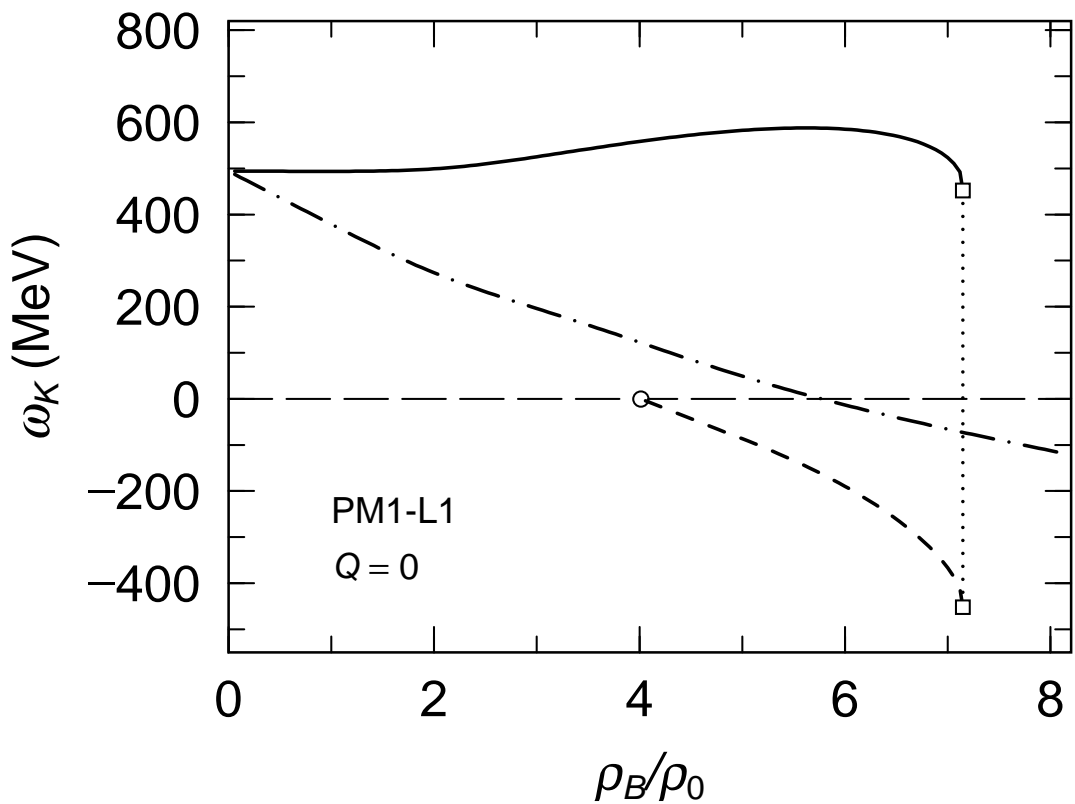}}
\caption
{\small 
Density-dependence of the kaon energy with zero momentum 
$\omega_K(0)$ with the parameter-set PM1-L1.
The solid, chain-dotted and dashed lines represent results for 
$K$, ${\bar K}$ and ${\bar K}_s$ ($s = -1$), respectively.
}
\label{Keng}
\end{figure}

In Fig.~\ref{Keng}, we show the density dependence of $\omega_K$ 
at $Q=0$ in PM1-L1 parameter set, where the solid, chain-dotted
and dashed  lines represent 
$\omega_K$ of $K$, ${\bar K}$ and ${\bar K}_s$, respectively.

\subsection{Effects of finite kaon momentum  ($Q\neq 0$ case)  }
\label{subsec:pwave}

\begin{figure}[ht]
\hspace*{0.3cm}
{\includegraphics[scale=0.7]{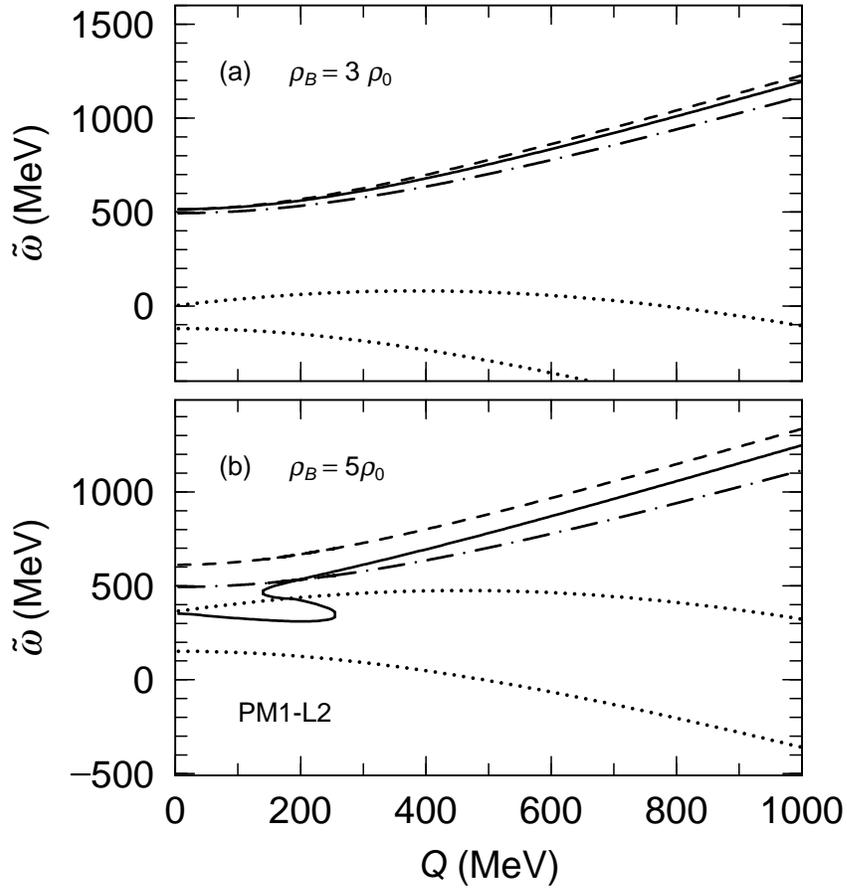}}
\caption{\small
The kaon dispersion relations with the PM1-L2 parameter set at 
$\rho_B = 3 \rho_0$ (a) and 5 $\rho_0$ (b).   
The solid and dashed lines show the kaon energies with and without the
 \LbN~loop, respectively. 
For reference, the energy of the free kaon is shown by the chain-dotted
 line.
The area surrounded by the two dotted lines indicates the continuum
 region where the kaonic states are unstable.
}
\label{KdisQ}
\end{figure}

\tpsp
In order to discuss the effects of the $p$-wave interaction coming from
the spatial component of 
the axial-vector interaction on the kaon self-energy, 
 we show, in Fig.~\ref{KdisQ}, the kaon dispersion relations at densities 
 $\rho_B$ = 3 $\rho_0$ (a) and 5 $\rho_0$ (b) for the PM1-L2 parameter set.
The solid and dashed lines indicate the kaon energies with and without
the \LbN~loop contribution, respectively.
For reference, we also plot the energy of the free kaon 
with the chain-dotted line. 
The area surrounded by the two dotted lines is the continuum region, where the kaonic modes are unstable.
One can immediately notice in Fig.~\ref{KdisQ} that the \LbN~loop 
contribution to the self-energy for the $K$ state is attractive. but the
total self-energy still gives a repulsive contribution for low densities
because of the TW term. 

In Fig.~\ref{KdisQ}~(a), the density $\rho_B$ (=3.0 $\rho_0$) is 
just the critical density for the appearance of the $\bar K_s$ mode with
null momentum, but still 
lower than the critical density at which the collective $\bar K_s$ mode
with a finite momentum $Q$ appears on the upper boundary of the continuum
region, $\tilde\omega=\epsilon_N(p_F(N))-\epsilon_\Lambda(p_F(N)-Q)$ for finite $Q$. 
Thus the $\bar K_s$ mode exists only with $Q=0$ at this density.
In Fig.~\ref{KdisQ}~(b), on the other hand, the density  $\rho_B$ (=5.0
$\rho_0$) is larger than 
the critical density $\rho_c^{(2)}$ (= 4.9 $\rho_0$) at which the normal
kaon state and ${\bar K}_s$ state disappear with $Q=0$. 
These two modes $K$ and $\bar K_s$ actually appear with a finite 
momentum $Q$ at this density. Note that the branches 
inside the continuum region have no physical meaning, because they are
not the real solutions of the dispersion equation, $D_K^{-1}(q_0;Q) =0$. 
At a certain value of $Q\sim 150$MeV, the $K$ and ${\bar K}_s$ 
come together just outside the continuum region, which may imply 
another instability with respect to $K \bar K_s$ condensation 
with a finite momentum. 
However, $K \bar K_s$ condensation actually proceeds at 
the lower critical density $\rho_c^{(2)}$ (=4.9 $\rho_0$) with $Q=0$. 

\begin{figure}[ht]
\hspace*{0.3cm}
{\includegraphics[scale=0.7]{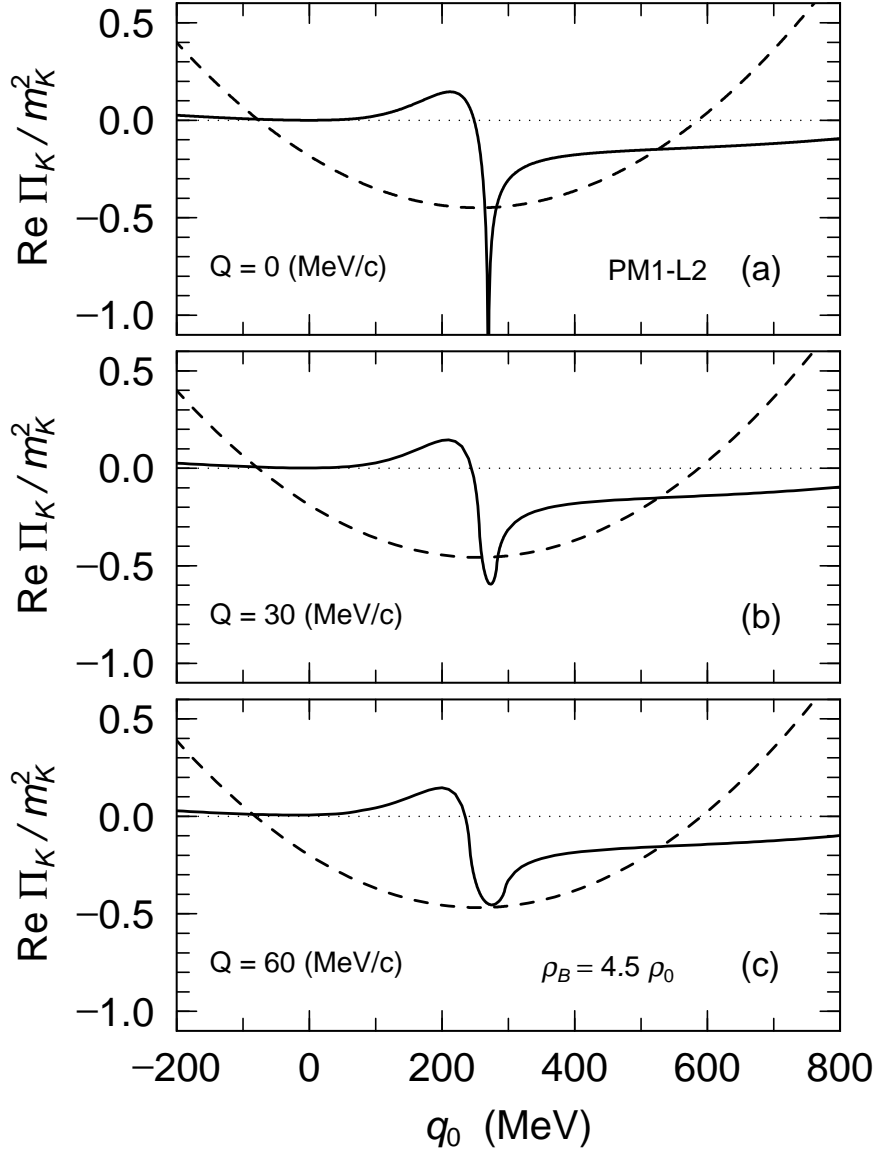}}
\caption{\small
The real part of the self-energy from the $\Lambda$-$N$~loop, 
${\rm Re} \Pi_K$ with the PM1-L2 parameter set at $\rho_B$= 4.5 $\rho_0 $ (the solid lines). 
The momentum is taken to be $Q = 0$ MeV/c for (a), 30 MeV/c  for (b), and 
60 MeV/c for (c). The dashed lines indicate $\Delta_s^{-1}(q_0;Q) $.
}
\label{KresQ}
\end{figure}

We should comment on why the results in the density region 
$\rho_c^{(1)} < \rho_B < \rho_c^{(2)}$ 
have not been shown in Fig.~\ref{KdisQ}. 
In reality, the collective $\bar K_s$ mode also exists  
with a small momentum outside the continuum region. 
However, because the minimum of ${\rm Re} \Pi_K(q_0;Q)$ has a finite depth, we cannot plot this dispersion relation clearly in the figure. 
Instead, in Fig.~\ref{KresQ}, we plot the contribution from the \LbN~loop
to the kaon self-energy ${\rm Re} \Pi_K(q_0;Q)$ at  $\rho_B$ = 4.5 $\rho_0 $ with the PM1-L2 parameter set in solid lines. The momentum is taken to be $Q = 0$ MeV/c for (a), 30 MeV/c for (b) and 60 MeV/c for  (c). The dashed lines indicate $\Delta_s^{-1}(q_0;Q)$. 
As the momentum $Q$ increases, starting with zero, the depth at the minimum of
${\rm Re} \Pi_K (q_0;Q)$ becomes shallower rapidly, and even the $\bar K_s$ branch enters  the continuum region, where both the ${\bar K}_s$ and $K_s$ are unphysical solutions to the dispersion equation, ${\rm Re} D_K^{-1}(q_0;Q)=0$. 
These apparent states disappear around $Q \approx 60$ (MeV/c) . 

 In general, as the momentum $Q$ increases ($Q \gtrsim$ 100 MeV/c), the minimum depth of the self-energy turns to be deeper again, and the position of the minimum shifts to a higher $q_0$. 
 At the same time, the width around the minimum becomes 
broad.\footnote{In the nonrelativistic framework, 
the Migdal function is usually used for the $p$-wave part of the 
particle-hole contribution to the meson 
self-energy\cite{kv03,bc79,ew88}. 
There is a qualitative agreement between the relativistic form of the 
particle-hole contribution to the self-energy and the nonrelativistic 
one with regard to the following features: (i)  the two extrema exist 
with negative and positive values within the boundaries of the 
continuum region. 
(ii) As the meson momentum increases, the minimum point shifts to a
 large $q_0$, the depth of the minimum increases as does  the width
 around the minimum. }
The collective $\bar K_s$ mode, if it should be formed, has a pole energy $\tilde \omega $ 
outside the upper boundary of the continuum region, where the $\Lambda
 N^{-1}$ self-energy is attractive, and the attraction becomes large as
 $Q$ increases for sufficiently large $Q$. However, this attraction  is
 not enough to overwhelm the increase of the kaon kinetic energy: In
 Fig.~\ref{KresQ}, even for a large $Q$($\gtrsim$ 60 MeV/c), we have no
 point of intersection of the solid curve with the dashed curve in the
 vicinity of the minimum.  After all,  the $\bar K_s$ state, once
 disappearing at $Q\sim$ 60 MeV/c, does not appear again for
 sufficiently large $Q$. Thus, it is concluded that, in symmetric
 nuclear matter, the onset of $K\bar K_s$ pair condensation is
 attributed to the $\Lambda$-$N$ loop contribution to the self-energy
 from the time component of the axial-vector interaction and that the
 $p$-wave interaction from the finite kaon momentum $Q$ does not assist
 the onset of $K\bar K_s$ condensation.\footnote{The sizable attractive
 contribution from the time component of the axial-vector interaction to
 the kaon self-energy has also been pointed out in case of neutron star
 matter within a relativistic framework in Ref.~\cite{kv03}. } 

This result, namely that the $p$-wave interaction given by the
finite $Q$ is not significant over the relevant densities 
in symmetric nuclear matter, is consistent with that obtained in
Ref.~\cite{m93}, where the role of the $p$-wave kaon-baryon 
interaction on kaon condensation was considered in neutron matter within a
nonrelativistic framework. 
It was shown that antikaon ($K^-$) condensation is realized
first from normal neutron matter at 
the lower density as a result of the $s$-wave scalar and vector attractive
interactions brought about by the $KN$ sigma term and the TW term, 
respectively. Furthermore, $p$-wave $K^-$ condensation starts at a higher
density and with finite momentum, accompanied by hyperon excitation, only after the development of the $s$-wave $K^-$condensed phase.  
In this respect, the $p$-wave kaon-baryon interaction is not 
responsible for the onset of kaon condensation realized from the normal phase.
\footnote{It should be noted that in Ref.~\cite{blr98} for example, 
the $\Sigma^-$ hyperon has been taken into account in their
consideration of  the in-medium kaon properties and that the role of 
''kaesobar '', a linear combination of the $K^-$ and $\Sigma^- N^{-1}$ 
particle-hole states through the $p$-wave interaction, has been
discussed. 
Such other hyperons may have an important effect on the behavior of 
the low-lying collective modes. 
In this paper, however, we only take into account the $\Lambda$ hyperon 
and clarify its effect on the kaon dispersion relation 
and onset mechanisms of $p$-wave kaon condensation within a simple situation. }

In Fig. \ref{KdisQ}, we have seen that the energy of the $K$ state 
is higher than the free kaon energy especially in the high momentum region.
This result shows us that the $K^+$ potential is totally repulsive.
It has been shown experimentally that only the elliptic flow 
of $K^+$ is out-of-plane, while other kinds of elliptic flow, 
such as nucleon and pion, occur in-plane, which is again consistent with 
the $K^+$ potential in high density matter being repulsive\cite{Keflow}. 
This experimental result is consistent with our theoretical result.
With regard to the possible experimental observation of  $K\bar K_s$ condensation,
 we can expect exotic processes caused by 
$K{\bar K}_s$ pair production in high-energy heavy-ion collisions. 
Such processes imply the enhancement of $\Lambda$ and $K$ production.  
In these processes, the final states must be $\Lambda$-$K$ pairs, 
 because ${\bar K}_s$ is a collective mode made from lambda-particle and nucleon-hole states, and most of the $\bar K_s$ modes are expected to be absorbed by the nucleons. 

\section{Discussions}

\subsection{Effects of the baryon masses}
\label{subsec:baryon}

\tpsp
In the preceding sections, we have seen an outstanding feature of the
$\Lambda$-$N$ loop contribution due to the time component of the axial-vector coupling term. 
On the other hand, the time component has been discarded in the usual
nonrelativistic framework, where the spatial component plays 
an important role\cite{m93,m02}. 
Hence it should be interesting to get deeper insight 
about the role of the time component of the axial-vector coupling term. 

One of the authors (T. Muto) has given the expression for the self-energy 
involving the $\Lambda$-$ N$ loop 
within the nonrelativistic framework \cite{m93,m02}. 
In that work, the static approximation in which baSSOEryons are 
at rest is used for the purpose of studying $p$-wave kaon condensation. 
In order to compare the present work with that,  
we apply the same static approximation, where the integrand of the 
$\Lambda N^{-1}$ self-energy,  Eq.~(\ref{eq:self}), 
is approximated to be the lowest-order term in the expansion with
$|\vp|/M_\alpha^\ast$. 
In the static approximation, we take the single particle energy to be 
$\epsi_\alpha (0) =  M_\alpha - U_s(\alpha) + U_0(\alpha)$ $(\alpha = N,
\Lambda)$, 
$\epsi_{\bar\Lambda} (0) =  M_{\Lambda} - U_s(\Lambda) - U_0(\Lambda)$.
Then, the kaon self-energy is given by:
\begin{equation}
\Pi_K (q_0;Q) = 
\frac{{\widetilde f} \ ^2}{2} \rho_B 
\Bigg\{ \frac{q_0^2}{\epsi_{N} (0) + \epsi_{\bar\Lambda} (0) - q_0  - i \delta}
+  \frac{Q^2}{\epsi_N (0) - \epsi_{\Lambda} (0) - q_0 - i \delta} \Bigg\} .
\label{static}
\end{equation}
The first term gives the repulsive $q_0^2$-dependent contribution from the
anti-lambda, but its magnitude is tiny. 
On the other hand, the second term gives
the attractive $Q^2$-dependent contribution, 
which is a major component within the static approximation. 
In Eq.(\ref{static}) one can immediately see that the $\Lambda$
particle contribution from the time component of 
the axial-vector interaction  [the first term on the r.~h.~s. of
Eq.~(\ref{eq:Res01}) ] does not exist 
because it is of order of $O(|\vp|^2/M_\alpha^{\ast 2})$ and 
is discarded in the static approximation. 
The absence of this term originates from the structure of the time component of the axial-vector coupling: 
\begin{equation}
\bar u_\Lambda ({\vec p}_\Lambda, s_\Lambda) \gamma_5 \gamma_0 
u_N ({\vec p}_N, s_N)
\approx \chi_{s_\Lambda}^\dagger 
(\frac{\vec{p}_\Lambda}{M_\Lambda^\ast} + \frac{\vec{p}_N}{M_N^\ast} ) \cdot
{\vec \sigma}
\chi_{s_N} ,
\label{NRtr}
\end{equation}
where ${\vec \sigma}$ is the Pauli matrix, 
$u_\alpha$ ($\alpha=\Lambda, N$) is a Dirac spinor, and
$\chi$ is the Pauli spinor. 
Apparently, the time component of the axial-vector
coupling gives no contribution in the static limit. 

Now we consider the term of order $O(|\vp|^2/M_{\alpha}^{\ast 2})$ 
in the kaon self-energy at $Q=0$.
Eq.~(\ref{eq:Res0}) can be rewritten as  
\begin{eqnarray}
\Pi_K (q_0; {\bf 0}) &= &\Pi_K^{(\Lambda)} (q_0; {\bf 0})+\Pi_K^{(\bar\Lambda)} (q_0; {\bf 0}) 
\nonumber \\
&=&-\frac{{\widetilde f}\ ^2}{2\pi^2} q_0^2
\int \frac{dp p^2\, n(\vp)}{E_\Lambda^\ast(\vp) E_N^* (\vp)} \Bigg\lbrack
\frac{E_\Lambda^\ast(\vp) E_N^\ast (\vp) +{\vp}^2-M_\Lambda^\ast M_N^\ast}
{E_\Lambda^\ast(\vp) - E_N^\ast (\vp)+q_0^\ast- i \delta } \nonumber \\
&+&\frac{-E_\Lambda^\ast(\vp) E_N^\ast (\vp) +{\vp}^2-M_\Lambda^\ast M_N^\ast}
{E_\Lambda^\ast(\vp) + E_N^\ast (\vp)-q_0^\ast + i \delta } \Bigg\rbrack \ , 
\label{eq:Res01}
\end{eqnarray}
where the first and second terms in the bracket on the r.h.s of
Eq.~(\ref{eq:Res01}) are the contributions from 
the $\Lambda$ and anti-$\Lambda$ excitations, respectively. 
Here we use the nonrelativistic description for the baryon energy 
as $E^*_{\alpha}(\vp)=M^*_{\alpha} + \vp^2/2M^*_\alpha$ 
$(\alpha = N, \Lambda)$, 
and neglect the anti-lambda contribution in Eq.~(\ref{eq:Res01}).
In this approximation the $\Lambda N^{-1}$ contribution to 
the kaon self-energy at $Q=0$ becomes  
\begin{equation}
\Pi_K (q_0; 0) = - \Big\lbrack\frac{\widetilde f}{2\pi}
\Big( \frac{ M_N^\ast + M_\Lambda^\ast}{ M_N^\ast M_\Lambda^\ast} \Big)
q_0\Big\rbrack^2 \int dp p^2 n(\vp)
\frac{\vp^2}
{\epsi^{nr}_\Lambda(\vp) - \epsi^{nr}_N(\vp) + q_0 -i\delta} \ , 
\label{eq:selfw}
\end{equation}
where the single particle energies are
$\epsi^{nr}_{\alpha}(\vp)= U_0(\alpha) + M^*_{\alpha} +
\vp^2/2M^*_\alpha$ $(\alpha = N, \Lambda)$.
This kaon self energy also has a divergent behavior as 
${\rm Re} \Pi_K (q_0;{\bf 0}) \sim - q_0^2 \log|q_0 + \deLN (p_F(N))|$,
and then the  collective mode (${\bar K}_s$) with $Q=0$ also appears 
in this nonrelativistic framework  in the restricted density
region,   $\rho_B=(2-3) \rho_0$,
where $q_0+\Delta\epsilon_{\Lambda N}(p_F(N))\sim 0$.
However, over the density $\rho_B\gtrsim 3\rho_0$, the contribution 
(\ref{eq:selfw}) becomes negligible, 
and the collective mode disappears (see the Appendix).

If the effective baryon masses are kept near the free masses,  
this nonrelativistic expansion is available in the density region 
$\rho_B \lesssim 6 \rho_0$.  
In the nonrelativistic framework for $p$-wave meson condensation, 
the effective baryon masses have usually been set to 
be free masses\cite{m93,m02,bc79,ew88}, and 
the kaon energy, $q_0$,  is small: $|q_0| <  Q$.
Therefore
the self-energy contribution coming from the time component 
of the axial-vector interaction has been neglected,
and the $p$-wave kaon condensation can be well described
with the static approximation.

However this argument is not always valid in different kinematical
conditions even in the nonrelativistic framework.
When the kaon momentum becomes very small, 
the major component given in the static approximation
becomes smaller,
and the minor component coming from the time component of the
axial-vector coupling becomes dominant 
(in the kinematical region  $|q_0| >> Q$).
This is why the collective mode caused by the $p$-wave 
interaction appears even in the zero kaon momentum limit 
at some density interval.

\begin{figure}[ht]
\hspace*{-0.3cm}
{\includegraphics[scale=0.75]{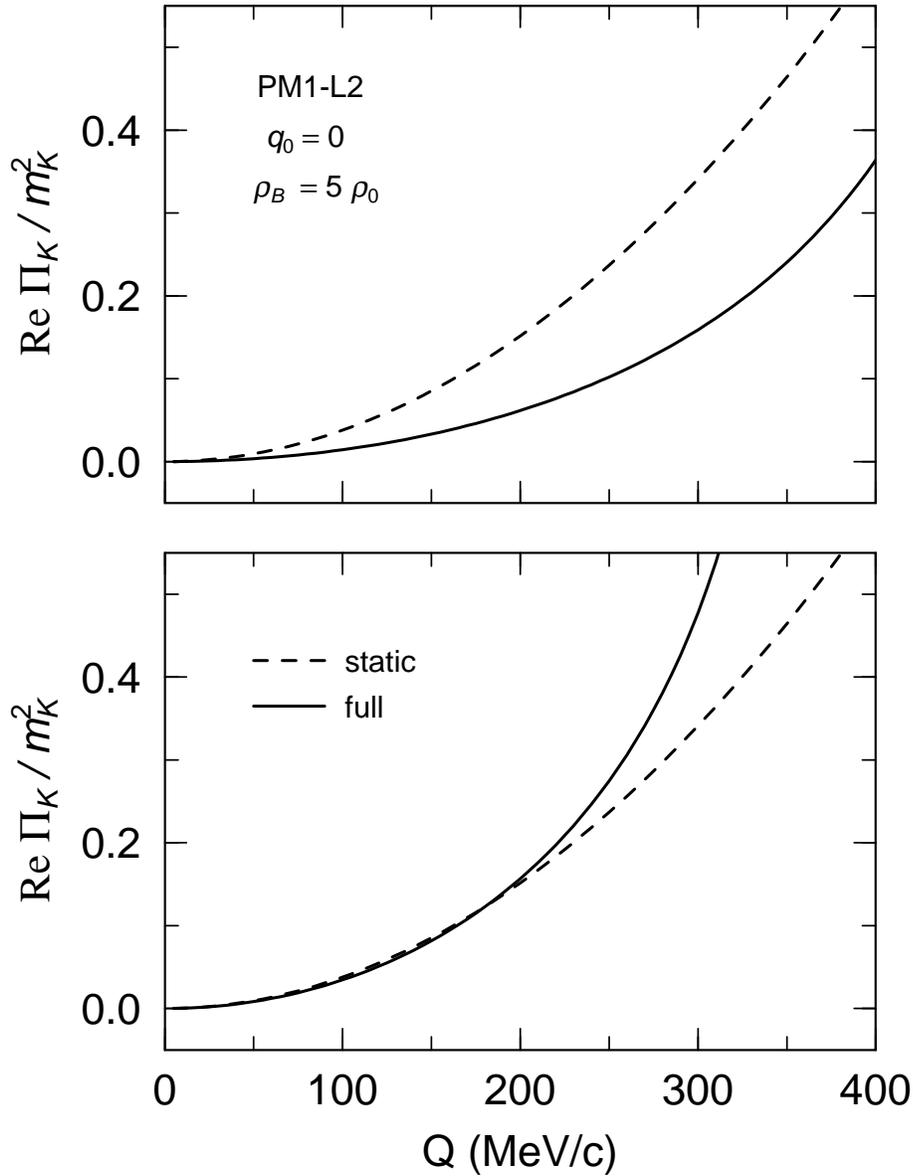}}
\caption{\small
The momentum dependence of the real part of the kaon self-energy 
at zero energy transfer  $q_0 = 0$, for 
$\rho_B$= 5.0 $ \rho_0$ using the parameter-set PM1-L2 (a), 
and no mean-fields (b). 
The solid and dashed lines represent the results in
the present approach 
(relativistic and non-static), and the static approach, respectively.}
\label{KrspQN}
\end{figure}

In the RMF theory 
the effective baryon masses decreases significantly, and 
the nonrelativistic expansion itself loses its validity. 
In Fig. \ref{KrspQN}, we plot the kaon self-energies at $q_0 = 0$ 
as functions of the kaon momentum, $Q$, at $\rho_B $= 5 $\rho_0$
in the non-static (solid line) and static calculations (dashed line)
using the PM1-L2 mean-fields (a) and no mean-fields (b).
We see that the static approximation cannot be used in the RMF approach, 
even in the finite momentum region $-$ though the static approximation
is available when the effective baryon masses are not too small. 
This can also be seen from the fact that  
the time component of the axial-vector current (\ref{NRtr}) becomes 
significant for the reduced effective nucleon mass. 
The importance of the time component of the axial-vector current in a many nucleon system (nuclear matter) is also well known 
as an explanation of the observed enhancement of the 0$^+$$-$0$^-$ 
beta transition in the lead region of finite nuclei\cite{w91}. 
The enhancement may be attributed to the decrease 
in the effective nucleon mass inside the nucleus. 

Thus the difference between our calculation and 
the static calculation is mainly seen around the kinetic region 
where the major component becomes zero.
The small effective masses enhance this difference.
This phenomenon also occurs in the energy denominator 
in the integrand of $\Pi_K$, $q_0 - \epsi_N + \epsi_\Lambda$.
In Eq.~(\ref{static}), the factor ${\rm Re} \Pi_K$ 
diverges at $q_0 = - \deLN(0)$,
though in the non-static calculation 
it diverges at $q_0 = - \deLN(p_F(N))$ when $Q=0$ 
and does not diverge when $Q$ is finite.
This difference is caused by the kinetic energies of the 
nucleon and lambda in the energy denominator.
Then $\Pi_K$ exhibits significant diffetrence between the static
and nonstatic calculations in the continuum region. 
In the study of the $p$-wave $K^{-}$ condensation \cite{m93,m02}
the collective energy is sufficiently far from the continuum region,
but we have to take care concerning this point in general studies. 

\subsection{Hyperon ($\Lambda$)-mixing effect }
\label{subsec:hyperon-mixing}

\tpsp
We have concentrated on the possiblity of kaon condensation 
in symmetric nuclear matter, where hyperons are not mixed and 
the total strangeness is constrained to be zero. 
In Sec.~\ref{subsec:pwave}, we have seen that the spatial component of
the axial-vector coupling is not responsible for the onset of kaon
condensation.  On the other hand, in Ref.~\cite{m02}, the role of the
$p$-wave kaon-baryon interaction with a finite momentum $Q$ has been
studied in {\it hyperonic matter}, 
where hyperons are mixed in the
ground state of neutron-star matter. 
It has been shown that the collective proton particle -$\Lambda$ 
hole ($p\Lambda^{-1}$) mode carrying the quantum number of $K$ (corresponding to the $K_s$) appears in a density regime where the $\Lambda$ fraction becomes larger than the proton fraction. 
At a certain density, $\bar K K_s$ condensation occurs with a finite momentum, stemming from the attractive $p$-wave kaon-baryon interaction. 

If  high-density matter, where many hyperons are mixed, is formed in hypernuclear experiments, 
kaon properties may also be modified through the nucleon particle-$\Lambda$ hole loop.  
Before elucidating the hyperon-mixing effect in the strangeness-conserving
case relativistically in future work, it is instructive to give here an outline of the onset mechanisms for $p$-wave kaon condensation in beta-equilibrated hyperonic matter, following the results of Ref.~\cite{m02}.  
We shall see that the effect of the spatial component of 
the axial-vector coupling may dominate over that 
of the time component in this case. 

First, for comparison, we recapitulate the result in Ref.~\cite{m93} for
the case where hyperons are not mixed 
in the ground state of neutron matter, and consider the possibility of
$K\bar K_s$ condensation only taking the $\Lambda$$N^{-1}$ loop into account for  hyperon excitations.  In the static approximation, the $p$-wave part of the $\Lambda N^{-1}$ self-energy is written with the help of Eq.~(\ref{static}) as 
\begin{equation}
\Pi_K ^{(\Lambda N^{-1})}(q_0;Q) =- \ 
\frac{{\widetilde f} \ ^2 Q^2\rho_N }{q_0 +\epsi_{\Lambda} (0) - \epsi_N (0) +i\delta} \ , 
\label{eq:nonrela-ln}
\end{equation}
where $\rho_N$ is the nucleon number density. The anti-lambda contribution  in Eq.~(\ref{static}) is omitted. The critical density for $K\bar K_s$ condensation is given by the double-pole condition, 
\begin{eqnarray}
& &{\rm Re} D_K^{-1}(q_0;Q)=0 \ ,\label{eq:dp1}\\
& & \partial {\rm Re} D_K^{-1}(q_0;Q) /\partial q_0=0 \ , \label{eq:dp2}
\end{eqnarray}
together with minimization of  ${\rm Re} D_K^{-1}(q_0;Q)$ with respect to $Q$, 
\begin{equation}
\partial {\rm Re} D_K^{-1}(q_0;Q) /\partial Q= -2 Q\Big\lbrack
 1-
\frac{{\widetilde f} \ ^2 \rho_N }{q_0 +\epsi_{\Lambda} (0) - \epsi_N (0) }\Big\rbrack =0 \ , 
\label{eq:dp3} 
\end{equation}
with the use of Eq.~(\ref{eq:nonrela-ln}). 
At a critical point for $K\bar K_s$ condensation, 
$q_0=-\omega_K(\bar K_s)=\omega_K(K)=O(m_K) $
($\sim$ 500 MeV), since the attractive interaction due to  the $KN$
sigma term is compensated by the repulsive vector interaction (the TW
term) for the kaon energy $\omega_K(K)$ and by the second-order effect
on the kaon self-energy\cite{fmmt96}.  
Also one can assume  $\epsi_{\Lambda} (0) - \epsi_N (0) >0$, since we
suppose the $\Lambda$ is not mixed into the ground state of neutron
matter. 
Then, the critical density for $K\bar K_s$ condensation with a finite
momentum is roughly estimated from Eq.~(\ref{eq:dp3}) as $\rho_c(K\bar
K_s)=(\omega_K(K)+\epsilon_\Lambda (0)-\epsilon_N(0))/\widetilde
f^2\gtrsim$ 9 $\rho_0$. As for $K\bar K_s$ condensation, the $p$-wave
interaction with a finite momentum is not effective enough to overwhelm
the kaon kinetic energy, and  the condition (\ref{eq:dp3}) is met only
for high densities.
This means that $K\bar K_s$ condensation in neutron matter is unlikely
to occur with a finite momentum $Q$ over the relevant
densities. Instead, $s$-wave $K^-$ condensation, induced by the sigma
and TW terms, proceeds in normal neutron matter\cite{m93}. 
This result should be compared with that of $p$-wave pion 
condensation, where the $p$-wave $\pi N$ attractive interaction 
is dominant in comparison with the $s$-wave interaction. 
In neutron matter, the $p$-wave part of the $\pi^+$ self-energy comes
from the proton particle-neutron hole excitation and is written in the
static approximation as $\displaystyle \Pi_\pi(q_0; Q)= 2(\tilde f_{\pi
NN} Q)^2\rho_n/q_0$, where $\tilde f_{\pi NN}=f_{\pi NN}/m_\pi$ with
$f_{\pi NN}(\simeq 1)$ being the $\pi N$ coupling constant and $m_\pi$
the pion mass. 
In the same way as $K\bar K_s$ condensation, the critical
density for $\pi^-\pi_s^+$ is estimated as
$\rho_c(\pi^-\pi_s^+)=-\omega_\pi(\pi_s^+) /(2\tilde f_{\pi NN}^2)\sim
\rho_0$, where
$\omega_\pi(\pi_s^+)=-\omega_\pi(\pi^-)$=$-O(m_\pi)$\cite{bc79,ew88}. 
As a result, $\pi^- \pi_s^+$ condensation occurs with a finite momentum at a rather low density. 

Now, we consider the case of hyperonic matter. 
For simplicity, we assume only the $\Lambda$'s are present as hyperons in neutron matter. Then both the $\Lambda N^{-1}$ and $N\Lambda^{-1}$ loops contribute to the kaon self-energy. 
 In the static approximation, one obtains 
\begin{eqnarray}
\Pi_K (q_0;Q) &=&\Pi_K^{(\Lambda N^{-1})} (q_0;Q) +\Pi_K^{(N \Lambda^{-1})} (q_0;Q)  \cr
&=&-
\frac{{\widetilde f} \ ^2 Q^2(\rho_N-\rho_\Lambda) }{q_0 +\epsi_{\Lambda} (0) - \epsi_N (0) +i\delta} 
\label{eq:nonrela-lnnl}
\end{eqnarray}
with $\rho_\Lambda$ being the $\Lambda$ number density. 
In addition to the double-pole condition, (\ref{eq:dp1}) and (\ref{eq:dp2}) with the $\Lambda N^{-1}$ and $N\Lambda^{-1}$ contributions to the kaon self-energy (\ref{eq:nonrela-lnnl}),  
the minimization condition for the kaon inverse propagator with respect to $Q$ is written as 
\begin{equation}
\partial {\rm Re} D_K^{-1}(q_0;Q) /\partial Q= -2 Q\Big\lbrack 1-\frac{{\widetilde f} \ ^2 (\rho_N-\rho_\Lambda) }{q_0 +\epsi_{\Lambda} (0) - \epsi_N (0) }\Big\rbrack =0 \ . 
\label{eq:dp3q}
\end{equation}
In case  of $K \bar K_s$ condensation, the pole energy at a critical point must be of $O(m_K)$ as seen above. 
 From the condition for chemical equilibrium, $\mu_\Lambda=\mu_N$, with respect to the weak process, $\Lambda\rightleftharpoons N\nu \bar\nu$, which is maintained in hyperonic matter, one can  estimate 
$ \epsi_{\Lambda} (0) - \epsi_N (0) =p_F(N)^2/(2M_N^\ast)-p_F(\Lambda)^2/(2M_\Lambda^\ast)=-O({\rm 10 \  MeV})$ with $p_F(\Lambda)$ being the Fermi momentum of  $\Lambda$. The denominator of the second term in the middle part of Eq.~(\ref{eq:dp3q}) 
is of $O(m_K)$, so that the $p$-wave contribution to the kaon self-energy cannot be comparable  with the kaon kinetic energy over the relevant densities. Thus, in this case, it is also difficult to meet the condition (\ref{eq:dp3q}) for $K \bar K_s$ condensation. 

On the other hand, in case of $\bar K K_s$ condensation, the double-pole condition renders $q_0=\omega_K(K_s)=-\omega_K(\bar K)$=$-O({\rm 10 \  MeV})$, since the $\bar K$ energy is significantly reduced from the free kaon mass due to the attractive $s$-wave scalar and vector interactions. 
 Together with the estimate that $\epsi_{\Lambda} (0) - \epsi_N (0) =-O({\rm 10 \  MeV})$, one can see that the $N\Lambda^{-1}$ contribution to the self-energy from the $p$-wave attractive interaction easily becomes large enough to satisfy the condition (\ref{eq:dp3q}) with a finite momentum $Q$, provided that the $\Lambda$ is much abundant than the nucleon, i.e., $\rho_\Lambda > \rho_N$. 
 At a critical point for $\bar K K_s$ condensation, one finds from Eq.~(\ref{eq:dp3q}) that 
 $\rho_N-\rho_\Lambda=(\omega_K(K_s)+\epsilon_\Lambda (0)-\epsilon_N(0))/\widetilde f^2$. For a numerical example, one obtains $\rho_c(\bar K K_s)$=3.7 $\rho_0$ at $Q$= 112 MeV/c and $\omega_K( K_s)$=$-$34 MeV for $\rho_\Lambda/\rho_B$=0.6 and $\rho_N/\rho_B$=0.4.\footnote{
The $\Lambda$-mixing ratio, $\rho_\Lambda/\rho_B$, in beta-equilibrated neutron star matter is not well known, because it depends on the specific models of the hyperon-nucleon and hyperon-hyperon interactions, 
in particular, at high densities\cite{ell,sch,hyp1,hyp2,hyp3}. } 
Hence, the $p$-wave kaon-baryon interaction from the spatial component 
of the axial-vector coupling plays an important role 
in the realization of  $\bar K K_s$ condensation in hyperonic matter. 

It is concluded within the nonrelativistic approach that 
the $p$-wave interaction coming from the spatial component 
of the axial-vector interaction is responsible for the onset and 
subsequent growth of the collective mode through 
the large $\Lambda$-mixing in the chemically-equilibrated matter. 
It is interesting to examine whether the consequences of the hyperon-mixing effect may also be applied commonly to the strangeness-conserving system within the RMF theory. 
The $K_s$ state carrying the same quantum number as the $K$ may appear 
as a low-lying collective mode generated through the $N\Lambda^{-1}$ loop. Then, there is a possibility of $\bar K K_s$ condensation, whose mechanism is similar to that of charged pion ($\pi^- \pi_s^+$) condensation\cite{bc79,ew88}. The strangeness-conserving system may also become unstable with respect to $\bar K_s$ condensation through the process, $\Lambda\rightarrow N+\bar K_s$, where the $\bar K_s$ state appears in the continuum region as an imaginary solution of the dispersion equation, $D_K^{-1}(q_0;Q)$=0. This  mechanism is similar to that of  $\pi^0$ condensation\cite{bc79,ew88}. 

\section{Summary and concluding remarks}

\tpsp
We have investigated the kaonic collective modes in high-density symmetric muclear matter 
by introducing the \LbN~loop as well as the usual $s$-wave
$KN$ interactions (the $KN$ sigma term and  the TW term) within the RMF theory. 
We have obtained two kinds of kaonic collective modes,
$K_s$ and ${\bar K}_s$ states, whose quantum numbers are the same as 
those of $K$ and ${\bar K}$, respectively.
One of these two collective modes, $K_s$, does not manifest as a physical
mode in zero strangeness nuclear matter 
because its pole energy always exists in the continuum region. 
 The appearance and growth of the ${\bar K}_s$ state is attributed to
 the time component of the axial-vector kaon-baryon interaction, which
 has a large contribution to the kaon  self-energy in the RMF theory,
 due to the fact that the effective baryon masses get very small for
 high baryon number densities. This result contrasts with the
 conventional nonrelativistic results on the $p$-wave pion/kaon
 condensations: In the nonrelativistic framework, the time component of
 the axial-vector interaction is of $O(p_F(N)^2/M_N^2)$, so
 that it has a  negligible effect on the meson self-energy. It has also
 been shown that the $p$-wave kaon-baryon interaction with a finite
 kaon momentum cannot assist the appearance of the $\bar K_s$ state in
 nuclear matter, because it is not attractive enough to overwhelm the energy increase 
 due to the meson kinetic energy term.  

Based on these results for the kaon properties in a medium, we have discussed the possibility of kaon condensation in high-density symmetric nuclear matter. 
We have found an instability of the system with respect to 
$K{\bar K_s}$ pair condensation. 
The driving force for $K\bar K_s$ condensation is brought about by the time component of the axial-vector kaon-baryon interaction, which works uniquely as a large attraction in the RMF theory. 
In this condensation, the $p$-wave interaction coming from the moving
nucleon, which is neglected in the usual static approxmation,
plays an important role.
On the other hand,  the moving kaon tends to hinder this condensation.
Then the $K{\bar K_s}$ pair condensation appears only in
the kinematical region where the kaon momentum is much smaller than
the absolute value of the kaon energy, $|q_0| >> Q$. 
The above argument is not special in RMF. 
The small effective mass in dense matter, however, increases
the baryon velocity and  enlarges effects from the nucleon Fermi motion. 

In this work, in-medium kaon properties have been considered in nuclear matter, where the lambda hyperon is not mixed in the ground state of matter. 
We may encounter another situation when we consider collective kaonic modes in matter where hyperons are largely mixed in the ground state.  In this case, the role of both the time and spatial components of the axial-vector interaction should be examined in detail.   
We also note that for the relativistic models considered here 
(which are similar to QHD) the decrease of $m_N$ with density 
is much stronger than in models of the QMC type \cite{tsht98,Bentz:2001vc}.
It will be very interesting to explore
the consequences of such models for the phenomenon of pair condensation
found here.

Finally, we comment on the possibility of kaon condensation in
high-energy heavy-ion collisions. 
The process we suggest in this work is consistent with the two
experimental results in SPS energy region, namely observation of
the elliptic flow \cite{Keflow} and the drastic enhancement
of the $K^+$ production \cite{adler}.
The elliptic flow shows  the repulsive $K^+$ potential 
in high density matter, and the drastic enhancement of $K^+$ production 
may indicate a large suppression of the $\Lambda$$K^+$ pair production 
energy there.
Of course there is still controversy at present 
about how the system reaches thermal equilibrium and 
to what extent the density and/or temperature are raised. 
Numerical simulations have shown that baryon density is in the range 
$\rho_B$= 7 $-$ 10 $\rho_0 $ can be achieved in the high-energy heavy-ion 
collisions at beam energies of several tens of GeV/u \cite{AGS1,AGS2}.
Experiments for our work must be available in
future facilities of J-PARC at JAERI/KEK and FAIR at GSI \cite{FAIR}. 
In this case, the system is expected to be quasi-equilibrium for a duration of about $4 - 8$ fm/c
with temperature $T \approx 120$MeV \cite{AGS1}, which is still below 
the pion mass. 
Hence, it is plausible that $K\bar K_s$ condensation may occur 
in the high-density regime.
However, in real experimental situations the system is strongly 
non-equilibrium, and a lot of resonant particles such as delta 
are produced in the compressed zone of the collisions. 
Thus, the model should eventually be extended to treat 
such non-equilibrium systems. 

Some transport models \cite{Cassing03,Mishra04} showed
that in-medium modification of kaon properties plays an important role
in heavy-ion collisions in 1.5 $-$ 2.0 GeV/u energy region, 
where the baryon density is estimated to be achieved to
be  only $2-4 \rho_0$ \cite{TOMO2}.
Then, heavy numerical simulations for several ten GeV/u 
must be performed in order to conclude
whether or not this phenomenon can be observed in heavy-ion collisions. 
One of the promising methods for this purpose is 
the RBUU approach \cite{LiKo,TOMO2,RBUU,Sahu1}. 
But in this case, 
we also need to introduce the momentum-dependence for the mean-fields 
in the high-energy region \cite{TOMO2,Sahu1}. 

In relation to the in-medium modification of kaon properties, the role of the $\Lambda$(1405)  as a $\bar K N$ bound state has been extensively studied\cite{k94,l98,wkw96,ro00,ske00,trp02}. 
In particular, self-consistent calculations of the $\bar K$ self-energy have been done with inclusion of both the Pauli-blocking for the nucleon and attraction coming from the modification of $K^-$ in the intermediate $\bar K N$ states. Subsequently, the effect of the $\bar K$ self-energy built on the $\bar K$ spectral density on the observables such as $K^-/K^+$ ratio in heavy-ion collisions has recently been discussed\cite{tprs03}. It is interesting to include the effects of the $\Lambda$(1405) and other subthreshold resonances in the $\bar K$ self-energy within our framework and to discuss their implications for heavy-ion collision experiments in future work. 

All discussions up to now are reasonable under the assumption
that the quark degrees of freedom do not affect the matter properties.
If the system goes to the quark-gluon-plasma (QGP) phase, we have to consider a different scenario. 
Even in such a case, kaon condensation may play an
important role for the strange particle production processes;
this phase may appear before or after the QGP phase stage. 
However, it is beyond the scope of the present study.

\vspace{2em}
\noindent{\bf Acknowledgments}

T.~Maruyama thanks the Institute for Nuclear Theory at University of
Washington for its hospitality.
The work of KT was supported by FAPESP contract 2003/06814-8 (Brazil). 
The work of T. Muto is supported in part by funds provided by 
Chiba Institute of Technology. 
The work of T.T. is partially supported by the Grant-in-Aid for the 21st Century COE
``Center for the Diversity and Universality in Physics '' from 
the Ministry of Education, Culture, Sports, Science and
Technology of Japan. It is also partially supported by the Japanese 
Grant-in-Aid for Scientific
Research Fund of the Ministry of Education, Culture, Sports, Science and
Technology (13640282, 16540246).
This work was also supported by DOE contract DE-AC05-84ER40150,
under which SURA operates Jefferson Laboratory. 

\appendix
\section*{Appendix}
\label{sec:appendix}

\tpsp
In order to figure out how the unphysical mode $K_s$ appears as a solution
to the dispersion equation ${\rm Re} D_K^{-1}(q_0; {\bf 0})$=0, we
consider the inverse propagator $D_K^{-1}(q_0; {\bf 0})$ in the complex $q_0$ plane.
For this purpose, however, the relativistic expression should not be
relevant  because the integration cannot be done analytically there.
Since, as mentioned in the text, the appearance of the $K_s$ and ${\bar
K}_s$ branches is 
independent of the relativistic approach,  
we discuss the $K_s$ pole by the use of the nonrelativistic
expression without loss of essential point in this Appendix.

The contribution from the $\Lambda$-$N$ loop to the kaon self-energy at $Q$=0 is given in the text as Eq.~(\ref{eq:selfw}).  
For simplicity, we take the free baryon masses instead of the effective masses 
and discard the baryon potentials $U(\alpha)$ ($\alpha=\Lambda, N$). 
Then the single-particle energy is written as 
$ \epsi_\alpha(\vp) = M_{\alpha} +\vp^2/(2 M_\alpha)$, and Eq.~(\ref{eq:selfw}) renders 
\begin{equation}
\Pi_K (q_0; Q=0) =
\Bigg\lbrack\frac{\widetilde f}{2\pi}
\Big( \frac{ M_N + M_\Lambda}{ M_N M_\Lambda} \Big)
\frac{q_0}{R}\Bigg\rbrack^2R p_F(N)^3\Big(\frac{1}{3}+z^2+\frac{1}{2}z^3\ln\frac{z-1}{z+1}\Big) \ , 
\label{eq:nonrelaself}
\end{equation}
where $R\equiv (M_\Lambda-M_N)/(2M_\Lambda M_N)$, 
and $z\equiv [( q_0+M_\Lambda-M_N)/R]^{1/2}/p_F(N)$.  
As seen from the integrand of Eq.~(\ref{eq:selfw}), the self-energy $\Pi_K(q_0; 0)$ 
has a cut in the complex $q_0$ plane with the interval,
$-\Delta\epsilon_{\Lambda N}(0)\leq q_0 \leq
-\Delta\epsilon_{\Lambda N}(p_F(N))$, i.e., 
$0\leq z\leq 1$, where  decay of the nucleon into the $\Lambda$ and the
kaon is kinematically allowed. 
Owing to the existence of the cut along the real axis, the imaginary part of the kaon
inverse propagator, 
${\rm Im} D_K^{-1}(q_0; 0)$, becomes discontinuous across the cut. 

\begin{figure}[ht]
\hspace*{-0.3cm}
{\includegraphics[scale=0.80]{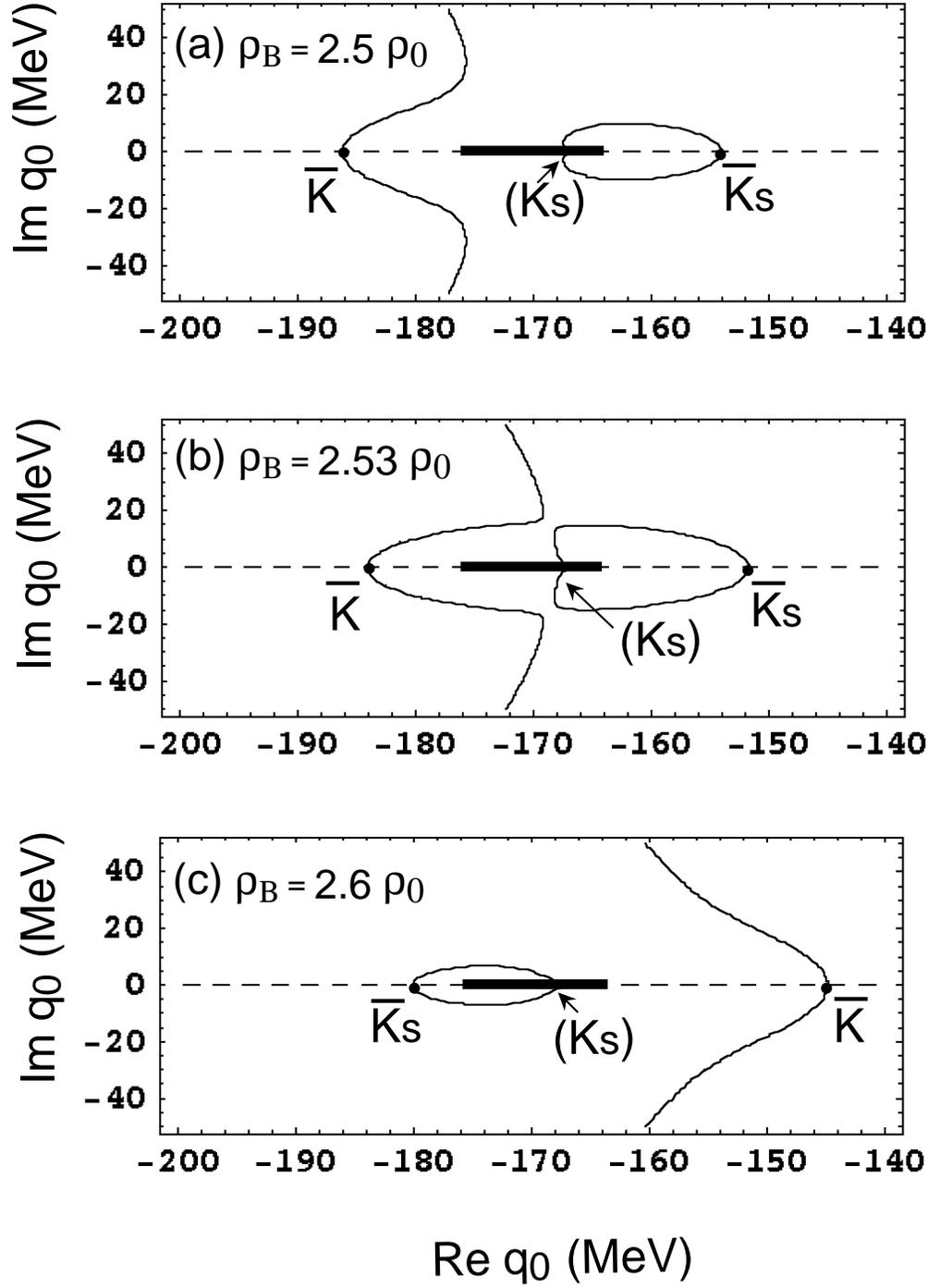}}
\caption{\small
The contour lines satisfying ${\rm Re} D_K^{-1}(q_0; 0)=0$ (the solid lines) and those satisfying ${\rm Im} D_K^{-1}(q_0; 0)=0$ (the dashed lines) on the complex $q_0$ plane at $\rho_B $= 2.5 $\rho_0$ for (a), 2.53 $\rho_0$ for (b), and 2.6 $\rho_0$ for (c). The nonrelativistic expression for the kaon self-energy is used.  
The bold bar along the real $q_0$ axis denotes the continuum region.
}
\label{fig:contour}
\end{figure}

The pole energies $\tilde \omega$ of the kaonic modes 
are obtained as the solutions to the dispersion equation, 
$ D_K^{-1}(q_0; 0)=0$, i.e., ${\rm Re} D_K^{-1}(q_0; 0)={\rm Im} D_K^{-1}(q_0; 0)=0$ 
in the complex $q_0$ plane. 
In Fig.~\ref{fig:contour}, we show the contour lines satisfying ${\rm Re} D_K^{-1}(q_0; 0)=0$ 
with the solid lines and those satisfying ${\rm Im} D_K^{-1}(q_0; 0)=0$ with the dashed lines 
at $\rho_B$=2.5 $\rho_0$ for (a), 2.53 $\rho_0$ for (b), and 2.6 $\rho_0$ for (c). 
The bold bar along the real $q_0$ axis denotes the continuum region. One finds that there are only the real solutions for the dispersion equation, corresponding to the $\bar K$, $\bar K_s$ and $K$ (The $K$ state is not depicted in the figure.). It should be noted that in the continuum region where ${\rm Im}
D_K^{-1}(q_0+i\delta; 0)\neq 0$, there appears 
an unphysical solution obtained only from the zero of the real part of the kaon inverse propagator, ${\rm Re} D_K^{-1}(q_0; 0) =0$. This solution corresponds to the $K_s$. 
However, this $K_s$ state does not really exist as a real solution to the dispersion equation, $D_K^{-1}(q_0; 0)=0$, nor even as a complex solution on the physical sheet of the $q_0$ plane. 

\begin{figure}[ht]
\hspace*{-0.3cm}
{\includegraphics[scale=0.80]{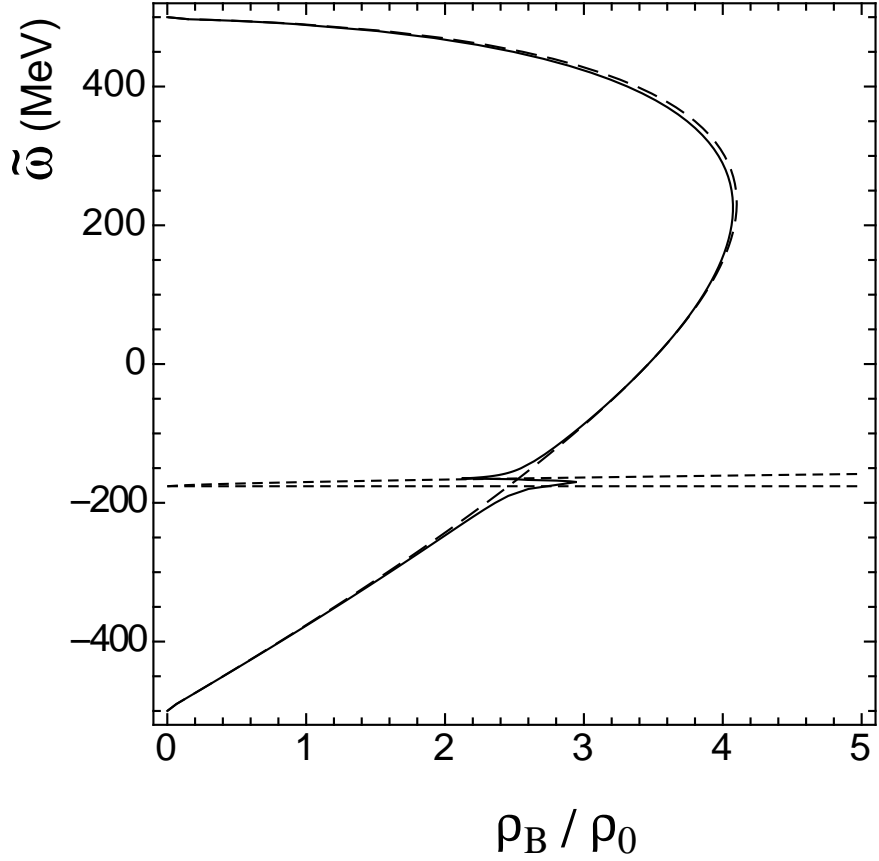}}
\caption{\small 
The density-dependence of the pole energies $\tilde\omega$ with the zero momentum transfer ($Q$=0), obtained from ${\rm Re} \ D_K^{-1}(q_0;{\bf 0})$=0 in the nonrelativistic limit. 
For comparison, those for which the $\Lambda N^{-1}$ contribution to the self-energy is put to be zero is shown by the dashed line. 
The dotted lines represent the two boundaries, $q_0=-\Delta\epsilon_{\Lambda N}(0)$ 
and $q_0=-\Delta\epsilon_{\Lambda N}(p_F(N))$, of the continuum region. }
\label{fig:w-rho}
\end{figure}

From Figs.~\ref{fig:contour}~(a)$-$(c), one can also see that the order of the $\bar K$ and $\bar K_s$ states is exchanged within a small density interval: Below the density $\rho_B$=2.53 $\rho_0$, the $\bar K$ state lies lower in pole energy than the $\bar K_s$ state [Fig.~\ref{fig:contour}~(a)]. Just above this density, the closed solid contour, which is relevant to the collective $\bar K_s$ and $K_s$ states, is connected outside the real axis with the open solid contour which is relevant to the $\bar K$ state.  With further increase in density, the $\bar K$ state is transferred to the location lying higher in pole energy than the $\bar K_s$ state [Fig.~\ref{fig:contour}~(c)] . Subsequently, the area of the closed contour gets small as density increases, and it disappears at a certain high density, where both the $K_s$ and $\bar K_s$ states disappear and only the $K$ and $\bar K$ states persist over the high densities. 
In Fig.~\ref{fig:w-rho}, we show the density-dependences of the pole energies $\tilde\omega$ with the zero momentum transfer ($Q$=0), obtained from the dispersion equation 
${\rm Re} \ D_K^{-1}(q_0;{\bf 0})$=0, by the solid line. For comparison, those for which the $\Lambda N^{-1}$ contribution to the self-energy is put to be zero is shown by the dashed line. 
The dotted lines represent the two boundaries, $q_0=-\Delta\epsilon_{\Lambda N}(0)$ 
and $q_0=-\Delta\epsilon_{\Lambda N}(p_F(N))$, of the continuum region. 
There appears an apparent branch of the $K_s$ mode in the continuum region 
for $\rho_B$=2$-$3 $\rho_0$. This unphysical state disappears together with the $\bar K_s$ state at $\rho_B\simeq$ 3 $\rho_0$.  
Therefore, the onset of pair condensation at $\rho_B$= 4.1 $\rho_0$ is
caused by the $K$ and $\bar K$ states in the nonrelativistic case, where
the  effective baryon masses are taken to be the free masses. It is to
be noted that the onset of  $K\bar K$  condensation is essentially
determined by the $s$-wave interactions brought about by the $KN$ sigma
term and the TW term and that the $\Lambda N^{-1}$ contribution to the
kaon self-energy given by Eq.~(\ref{eq:nonrelaself}) is hardly effective
around this critical density.

\newpage

\newpage

\newpage

\newpage

\newpage

\newpage

\vspace*{5mm}

\newpage

\newpage

\end{document}